\documentclass[superscriptaddress,a4paper,twocolumn]{revtex4-2}
\usepackage{geometry}
\usepackage[english]{babel}
\usepackage[latin1]{inputenc}
\usepackage[hidelinks]{hyperref}
\usepackage{amsfonts}
\usepackage{amsmath, amssymb, amsfonts, mathrsfs}
\usepackage{graphicx}
\usepackage{xspace}
\usepackage{multirow}
\usepackage{array} 
\usepackage{mathtools}

\usepackage{float}
\usepackage{bbold}

\newcolumntype{P}[1]{>{\centering\arraybackslash}p{#1}}
\newcolumntype{M}[1]{>{\centering\arraybackslash}m{#1}}

\usepackage[bottom]{footmisc}

\newcommand{\ket}[1]{|#1\rangle}
\newcommand{\bra}[1]{ \langle #1 \,|}
\newcommand{\mean}[1]{\langle #1 \rangle}

\makeatletter
\renewcommand{\fnum@figure}{Fig. \thefigure}
\makeatother

\begin{document}

\title{Experimental full network nonlocality with independent sources and strict locality constraints}

\author{Xue-Mei Gu}
\thanks{These authors contributed equally to this work.}
\author{Liang Huang}
\thanks{These authors contributed equally to this work.}

\affiliation{Hefei National Research Center for Physical Sciences at the Microscale and School of Physical Sciences, University of Science and Technology of China, Hefei 230026, China}
\affiliation{Shanghai Research Center for Quantum Science and CAS Center for Excellence in Quantum Information and Quantum Physics, University of Science and Technology of China, Shanghai 201315, China}
\affiliation{Hefei National Laboratory, University of Science and Technology of China, Hefei 230088, China}

\author{Alejandro Pozas-Kerstjens}
\thanks{These authors contributed equally to this work.}
\affiliation{Instituto de Ciencias Matem\'aticas (CSIC-UAM-UC3M-UCM), 28049 Madrid, Spain}
\affiliation{Departamento de An\'alisis Matem\'atico, Universidad Complutense de Madrid, 28040 Madrid, Spain}

\author{Yang-Fan Jiang}
\affiliation{Jinan Institute of Quantum Technology, Jinan 250101, China}

\author{Dian Wu}
\author{Bing Bai}
\author{Qi-Chao Sun}
\author{Ming-Cheng Chen}
\author{Jun Zhang}
\affiliation{Hefei National Research Center for Physical Sciences at the Microscale and School of Physical Sciences, University of Science and Technology of China, Hefei 230026, China}
\affiliation{Shanghai Research Center for Quantum Science and CAS Center for Excellence in Quantum Information and Quantum Physics, University of Science and Technology of China, Shanghai 201315, China}
\affiliation{Hefei National Laboratory, University of Science and Technology of China, Hefei 230088, China}

\author{Sixia Yu}
\affiliation{Hefei National Research Center for Physical Sciences at the Microscale and School of Physical Sciences, University of Science and Technology of China, Hefei 230026, China}

\author{Qiang Zhang}
\author{Chao-Yang Lu}
\author{Jian-Wei Pan}
\affiliation{Hefei National Research Center for Physical Sciences at the Microscale and School of Physical Sciences, University of Science and Technology of China, Hefei 230026, China}
\affiliation{Shanghai Research Center for Quantum Science and CAS Center for Excellence in Quantum Information and Quantum Physics, University of Science and Technology of China, Shanghai 201315, China}
\affiliation{Hefei National Laboratory, University of Science and Technology of China, Hefei 230088, China}

\begin{abstract}
Nonlocality arising in networks composed of several independent sources gives rise to phenomena radically different from that in standard Bell scenarios. Over the years, the phenomenon of network nonlocality in the entanglement-swapping scenario has been well investigated and demonstrated. However, it is known that violations of the so-called \textit{bilocality inequality} used in previous experimental demonstrations cannot be used to certify the non-classicality of their sources. This has put forward a stronger concept for nonlocality in networks, called full network nonlocality. Here, we experimentally observe full network nonlocal correlations in a network where the source-independence, locality, and measurement-independence loopholes are closed. This is ensured by employing two independent sources, rapid setting generation, and space-like separations of relevant events. Our experiment violates known inequalities characterizing non-full network nonlocal correlations by over five standard deviations, certifying the absence of classical sources in the realization.
\end{abstract}

\maketitle

Bell's theorem \cite{bell1964einstein}, stating that quantum predictions are incompatible with local realism, has deeply influenced our understanding of physics. Specifically, the correlations obtained by local measurements on a remotely shared quantum system cannot be explained by a local hidden variable (LHV) model. This is generally known as Bell nonlocality \cite{brunner2014bell}, which has been confirmed in numerous Bell experiments \cite{shalm2015strong,giustina2015significant,hensen2015loophole,rosenfeld2017event,li2018test} via violations of Bell inequalities \cite{bell1964einstein,clauser1969proposed}. Apart from its fundamental interest, Bell nonlocality has also found numerous applications as an indispensable resource in device-independent quantum information tasks \cite{acin2007device,acin2016certified,pironio2010random}.

\begin{figure}[!t]
	\centering
	\includegraphics[width=0.46\textwidth]{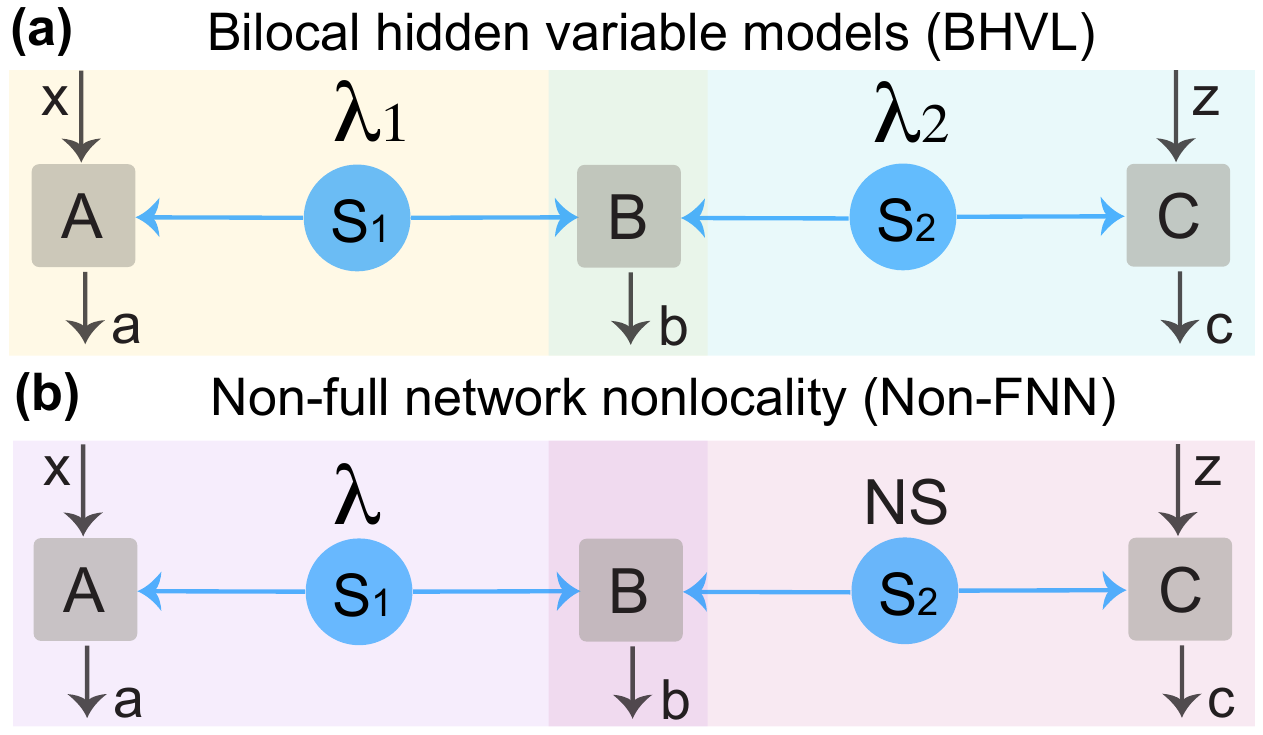}
	\caption{A network with sources (blue circles) distributing physical systems (blue arrows) to three nodes Alice, Bob and Charlie, represented as A, B, and C (gray squares). ($x,a$) and ($z,c$) are Alice's and Charlie's inputs and outputs. Bob performs a joint measurement on his systems yielding outcomes $b$. \textbf{(a)} BLHV model, where each source distributes a different LHV ($\lambda_{1}$ and $\lambda_{2}$). \textbf{(b)} Test for FNN, where the model to be discarded consists of one LHV $\lambda$ and a general nonlocal source (NS).} 
	\label{Fig:concepts}
\end{figure}

In the standard Bell tests for LHV models, local influences are mediated by a single, common LHV that is shared among all the parties. Recently, growing interest has been devoted to the exploration of Bell nonlocality in networks (see \cite{tavakoli2021bell} for a recent review). The simplest example is the entanglement-swapping scenario \cite{pan1998experimental}, where two independent parties that are not causally connected can become entangled. While correlations generated in networks can be contrasted against standard LHV models, it is more natural and physically motivated to consider models with independent hidden variables that reproduce the network structure. In the entanglement-swapping network, this gives rise to the study of bilocal hidden variable (BLHV) models and the associated phenomenon of (non)-bilocality \cite{branciard2010characterizing, branciard2012bilocal}. Importantly, recent experiments \cite{carvacho2017experimental, saunders2017experimental, andreoli2017experimental, sun2019experimental, poderini2020experimental} show that there are correlations that admit a standard LHV model but nonetheless are incompatible with a BLHV model (depicted in Fig. \ref{Fig:concepts}(a)).

It was recently argued that the violations of the inequalities characterizing correlations with BLHV models did not capture all the intricacies of nonlocality in the entanglement-swapping network \cite{pozas2022full}. Concretely, the bilocality inequality in \cite{branciard2012bilocal} can be violated when just one of the sources is entangled~\cite{saunders2017experimental}. More importantly, all its quantum violations can be simulated with a strategy of the form depicted in Fig. \ref{Fig:concepts}(b), this is, keeping one classical source if the other one is allowed to distribute systems only limited by the no-signaling principle (e.g., a Popescu-Rohrlich box \cite{popescu1994quantum} that allows stronger-than-quantum correlations) \cite{pozas2022full}. These systems are currently a hypothetical construct, but it is useful to consider them in order to be able to make statements independent of the theory describing the physical systems involved \cite{gisin2020constraints,coiteuxroy2021}.

In other words, a violation of the bilocality inequality only guarantees that at least one of the sources in the network is non-classical. In order to have a device-independent certification that both sources are non-classical, one must (i) insert the additional hypothesis that quantum mechanics describes nature at its fundamental level and (ii) achieve a violation of the bilocality inequality beyond the value described in \cite{saunders2017experimental}.
But if the additional assumption (i) is removed, no value of the inequality guarantees the absence of classical sources in the network.

This put forward a stronger, arguably more genuine, definition of network nonlocality, where distributions regarded as interesting are those that cannot be explained by having at least one classical source, regardless of the rest. This is known as full network nonlocality (FNN) \cite{pozas2022full}, and its observation implies that all the sources used in the network must necessarily uphold some degree of non-classicality.
FNN has been since then observed in experimental realizations of the entanglement-swapping scenario featuring three- \cite{haakansson2022experimental} and four-outcome \cite{huang2022experimental} measurements in the central party, and in the three-branch star network \cite{wang2022experimental}. Importantly, the independence of the sources in these experiments is assumed rather than enforced. This opens the possibility to the existence of an LHV simulation of the correlations observed \cite{lee2018towards}, known as the \emph{source independence loophole}. Similarly, the independence of the measurement choices and the isolation of the parties are also assumed, leading to the \emph{freedom of choice} and the \emph{locality} loopholes, respectively.

Here, we report the first experimental demonstration of the existence of FNN in quantum theory in a realization where the network structure is strictly guaranteed. This is, we show that there exist probability distributions generated in quantum networks that cannot be simulated if one of the sources distributes classical systems, even if the rest are allowed to distribute (so far, hypothetical) stronger-than-quantum systems. We do so by violating the FNN witnesses of \cite{pozas2022full} in an optical network that distributes quantum systems generated from independent sources under strict locality conditions, i.e., in which all the parties involved are space-like separated.

\paragraph*{FNN in the entanglement-swapping network.} FNN correlations are defined, analogously to standard nonlocal ones, as not admitting a specific model. In the entanglement-swapping scenario, this model is captured in Fig. \ref{Fig:concepts}(b). The correlations generated when the source $\text{S}_1$ is classical (i.e., an LHV) are described by $p(a,b,c|x,z)=\int d\lambda\rho(\lambda)p(a|x,\lambda)p(b,c|\lambda,z)$,
where $\rho(\lambda)$ is the probability distribution characterizing the LHV $\lambda$ between Alice and Bob, $p(a|x,\lambda)$ is Alice's response function, and $p(b,c|\lambda,z)$ is a joint response function for Bob and Charlie, which in general is only constrained by no-signaling (i.e., by $\sum_c p(b,c|\lambda,z)=p(b|\lambda)$ and $\sum_b p(b,c|\lambda,z)=p(c|z)$). To establish FNN, one needs to consider also the case with interchanged sources. Ref. \cite{pozas2022full} showed that, in the entanglement-swapping scenario when $a,c,z,x\in\{0,1\}$ and $b\in\{0,1,2\}$, the simultaneous violation of both the following inequalities certifies FNN:
\begin{equation}
	\begin{aligned}
		\mathcal{R}_{\text{C-NS}}:= & 2\left\langle A_{0} B_{1} \left( C_{0} - C_{1} \right) \right\rangle + \left\langle A_{1} B_{0} \left( 2C_{0} + C_{1} \right) \right\rangle\\
		& - \left\langle B_{0} \right\rangle + \left(\left\langle A_{1} B_{0} \right\rangle + \left\langle B_{0} C_{0} \right\rangle - \left\langle C_{0} \right\rangle \right) \left\langle C_{1} \right\rangle \leq 3,
	\end{aligned}
	\label{eq:functionF1}
\end{equation}
and \begin{equation}
	\begin{aligned}
		\mathcal{R}_{\text{NS-C}}:= & 2\left\langle A_{0} B_{1} \left( C_{0} - C_{1} \right) \right\rangle + \left\langle A_{1} B_{0} \left( C_{0} + 2 C_{1} \right) \right\rangle  \\
		& - \left\langle B_{0} \right\rangle + \left\langle A_{1} \right\rangle (\left\langle A_{1} B_{0} \right\rangle + \left\langle B_{0} C_{1} \right\rangle \\
		& + \left\langle C_{0} - C_{1} - A_{1} \right\rangle)\leq 3,
	\end{aligned}
	\label{eq:functionF2}
\end{equation}
where expectation values are computed following \cite{branciard2012bilocal}, namely $\mean{A_x B_0 C_z} =\sum_{a,b,c}(-1)^{a+c+[b>1]}p(a,b,c|x,z)$ and $\mean{A_x B_1 C_z} = \sum_{a,b\in\{0,1\},c}(-1)^{a+b+c}p(a,b,c|x,z)$, the function $[p]$ evaluating to $0$ if $p$ is true and to $1$ otherwise (see Appendix \ref{app:analytic} for further details).

Importantly, these inequalities can be violated simultaneously in quantum networks \cite[App. F]{pozas2022full}. Take both sources to emit a Bell state $\ket{\Phi^{+}}$, Alice and Charlie to perform measurements $A_{x}\in\{X,Z\}$ and $C_{z}\in\{\frac{Z+X}{\sqrt{2}},\frac{Z-X}{\sqrt{2}}\}$, where $X$ and $Z$ are the Pauli operators, and Bob to perform a partial Bell state measurement (BSM) with three outputs $\{\Phi^{+}, \Phi^-, \mathbb{1}-\Phi^{+}-\Phi^-\}$ that are correspondingly associated to the outcomes $b\in\{0,1,2\}$, where $\Phi^{\pm}=\ket{\Phi^{\pm}}\bra{\Phi^{\pm}}$. The resulting distribution $p(a,b,c|x,z)$ leads to the violations $\mathcal{R}_{\text{C-NS}}=\mathcal{R}_{\text{NS-C}}=5/\sqrt{2}\approx3.5355$.

\begin{figure*}[!t]
	\centering
	\includegraphics[width=0.99\textwidth]{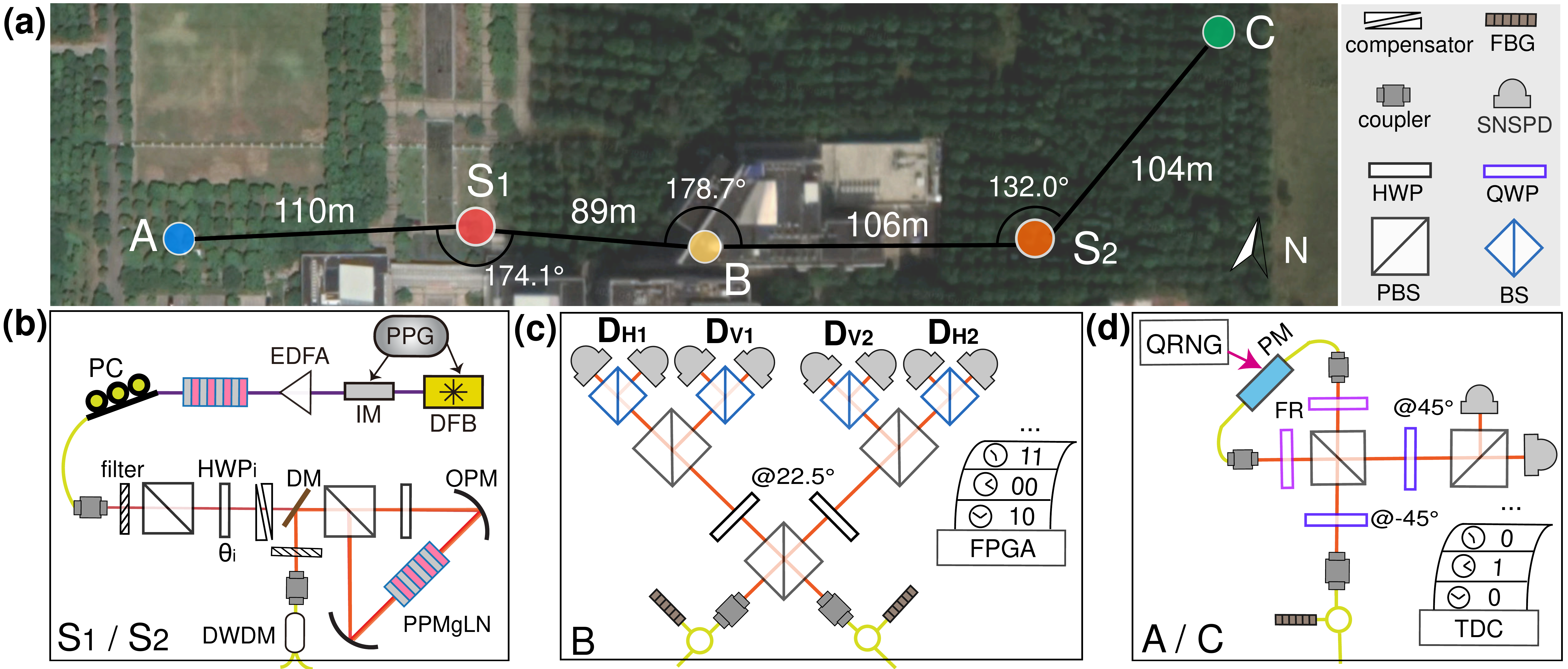}
	\caption{Experimental realization. \textbf{(a)} Overview of our quantum network with two independent sources ($\text{S}_{1}$ and $\text{S}_{2}$) and three nodes (Alice, Bob, and Charlie, represented as A, B, and C). Geographic picture is taken from Google Maps, $\copyright$2022 Google. \textbf{(b)} By pumping a periodically poled MgO doped lithium niobate (PPMgLN) crystal in a Sagnac loop, photon pairs in the state $\ket{\varphi}_{i}=\cos{2\theta_{i}}\ket{HH}+\sin{2\theta_{i}}\ket{VV}$ are created via SPDC in each source, where $\theta_{i}$ is the angle of $\text{HWP}_{i}$ mounted on a motorized rotator. The pulse pattern generator (PPG) in $\text{S}_{1}$ acts as a master clock for synchronizing all devices (see Refs.~\cite{sun2019experimental,wu2022experimental} for details). \textbf{(c)} Bob performs a partial Bell state measurement \cite{pan1998greenberger} and his photon detection events are real-time analyzed and recorded by a field-programmable gate array (FPGA). \textbf{(d)} Alice and Charlie measure their photons according to the inputs from their private quantum random number generators (QRNGs). Their detection events and setting choices are recorded by a time-to-digital converter (TDC). See the appendices for more details. IM, intensity modulator; EDFA, erbium-doped fiber amplifier; PC, polarization controller; DM, dichroic mirror; OPM, off-axis parabolic mirror; DWDM, dense wavelength-division multiplexer; FBG, fiber Bragg grating; PBS, polarizing beam splitter; HWP, half-wave plate; QWP, quarter-wave plate; BS, beam splitter; SNSPD, superconducting nanowire single photon detector; FR, Faraday rotator; PM, electro-optic phase modulator.} 
	\label{Fig:setups}
\end{figure*}

\paragraph*{Experimental realization.} We implement this in a photonic network, illustrated in Fig. \ref{Fig:setups}(a). Two independent sources $\text{S}_{1}$ and $\text{S}_{2}$ distribute entangled photons to three separate observers Alice, Bob, and Charlie. Each source generates a polarization-entangled state $\ket{\varphi}_{i}=\cos{2\theta_{i}}\ket{HH}+\sin{2\theta_{i}}\ket{VV}$ via type-0 spontaneous parametric down-conversion (SPDC), as shown in Fig. \ref{Fig:setups}(b). In each source, a pulse pattern generator (PPG) sends 250 MHz trigger signals to drive a distributed feedback (DFB) laser such that its electric current switches from much below to well above the lasing threshold, indicating that the phase of each generated pump pulse is randomized \cite{abellan2015generation}. In this way, we erase any quantum coherence between the pump pulses and disconnect the two SPDC processes on each experimental trial, thus closing the source-independence loophole under the reasonable assumption that hidden variables are created together with state emission (further details see Refs.~\cite{sun2019experimental,wu2022experimental}).

At the central node, Bob sandwiches polarization beam splitters (PBSs) between two half-wave plates (HWP@22.5$^{\circ}$) to realize a partial BSM \cite{pan1998greenberger} that distinguishes Bell states $\ket{\Phi^+}$, $\ket{\Phi^-}$, and a remaining group of Bell states $\{\ket{\Psi^+}$, $\ket{\Psi^-}\}$ (which is $\mathbb{1}-\Phi^+-\Phi^-$) by the coincidence detection among the pseudo-number-resolving detectors depicted as $\text{D}_{\text{H}1}$, $\text{D}_{\text{V}1}$, $\text{D}_{\text{H}2}$, and $\text{D}_{\text{V}2}$ in Fig. \ref{Fig:setups}(c). Bob's photon detections are analyzed in real time and recorded by a field-programmable gate array. Once Bob obtains a BSM output, he sends the corresponding timestamps information to Alice and Charlie. We implement a high-speed, high-fidelity, single-photon polarization analyzer at a rate of 250 MHz at Alice's and Charlie's location (Fig. \ref{Fig:setups}(d)), where the measurement choice depends on random bits produced from private fast quantum random number generators (QRNGs), see Appendix \ref{app:qrng}. All random bits from the QRNGs pass the NIST randomness tests \cite{rukhin2010statistical} (for more details, see Refs.~\cite{liu2018device,sun2019experimental,wu2022experimental}). All setting results and detections are locally recorded by their time-to-digital converters that are fed with Bob's timestamps information. All locally stored data are collected by a separate computer, in which we post-select the four-photon coincidence events by aligning all to Bob's BSM timestamps and use the four-photon coincidences for $\mathcal{R}_{\text{C-NS}}$ and $\mathcal{R}_{\text{NS-C}}$.

\begin{figure*}[!t]
	\centering
	\includegraphics[width=1\textwidth]{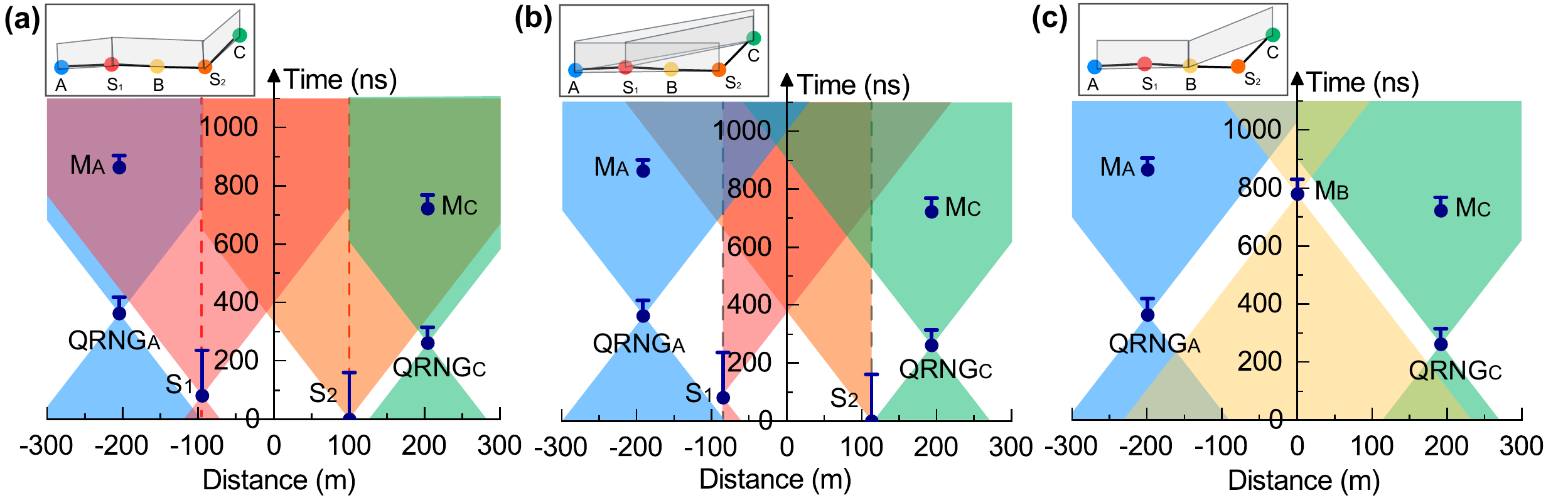}
	\caption{Space-time configuration in each experimental trial. \textbf{(a)}-\textbf{(c)} Space-time diagrams of the relationship between important events in the nodes, as displayed in the boxes. Blue vertical bars in each space-time diagram denote the time elapsing for the relevant events, with start and end marked by circles and horizontal lines, respectively. \textbf{(a)} Left, middle, and right panels split by red dashed lines: space-time analysis between $\text{QRNG}_{\text{A}}$ and $\text{S}_{1}$, space-time analysis between $\text{S}_{1}$ and $\text{S}_{2}$, and space-time analysis between $\text{S}_{2}$ and $\text{QRNG}_{\text{C}}$. \textbf{(b)} Space-time analysis between $\text{QRNG}_{\text{A}}$ and $\text{QRNG}_{\text{C}}$, and space-like separations between $\text{QRNG}_{\text{A}}$ ($\text{QRNG}_{\text{C}}$) and $\text{M}_{\text{C}}$ ($\text{M}_{\text{A}}$), and between $\text{S}_{1}$ ($\text{S}_{2}$) and $\text{M}_{\text{C}}$ ($\text{M}_{\text{A}}$). \textbf{(c)} Space-time analysis between $\text{M}_{\text{B}}$ and $\text{QRNG}_{\text{A}}$ ($\text{QRNG}_{\text{C}}$). All the space-time relations are drawn to scale. For more details see Appendix \ref{app:spacetime}.}
	\label{Fig:spacetime}
\end{figure*} 
We confirm the space-time configuration and characterize the delays of all relevant events, namely: (1) emission ($\text{S}_{1}$ and $\text{S}_{2}$), i.e., the photon creation events in the sources; (2) the choice of measurement setting ($\text{QRNG}_{\text{A}}$ and $\text{QRNG}_{\text{C}}$), i.e., completing the quantum random number generation at Alice's and Charlie's nodes for choosing their measurements; and (3) the measurement ($\text{M}_{\text{A}}$, $\text{M}_{\text{B}}$, and $\text{M}_{\text{C}}$), i.e., finishing the single photon detection by Alice, Bob, and Charlie. In the experiment, the time reference to synchronize all events is set to the 12.5 GHz internal microwave clock of the PPG in source $\text{S}_{1}$ \cite{wu2022experimental}. By employing QRNG and space-like separating setting choice and measurement on one side from the measurement on other sides, we close the locality loophole. By space-like separating setting choice events from state creation events (which is also the origin of a hidden variable), we also close the measurement-independence loophole. Finally, the emission at $S_1$ is spacelike separated from Charlie's measurement, and analogously for Alice's measurement and the emission at $S_2$. The details about the space-like separation of all relevant events are shown in Fig.~\ref{Fig:spacetime} and Appendix \ref{app:spacetime}.

\begin{figure}[!t]
	\centering
	\includegraphics[width=0.5\textwidth]{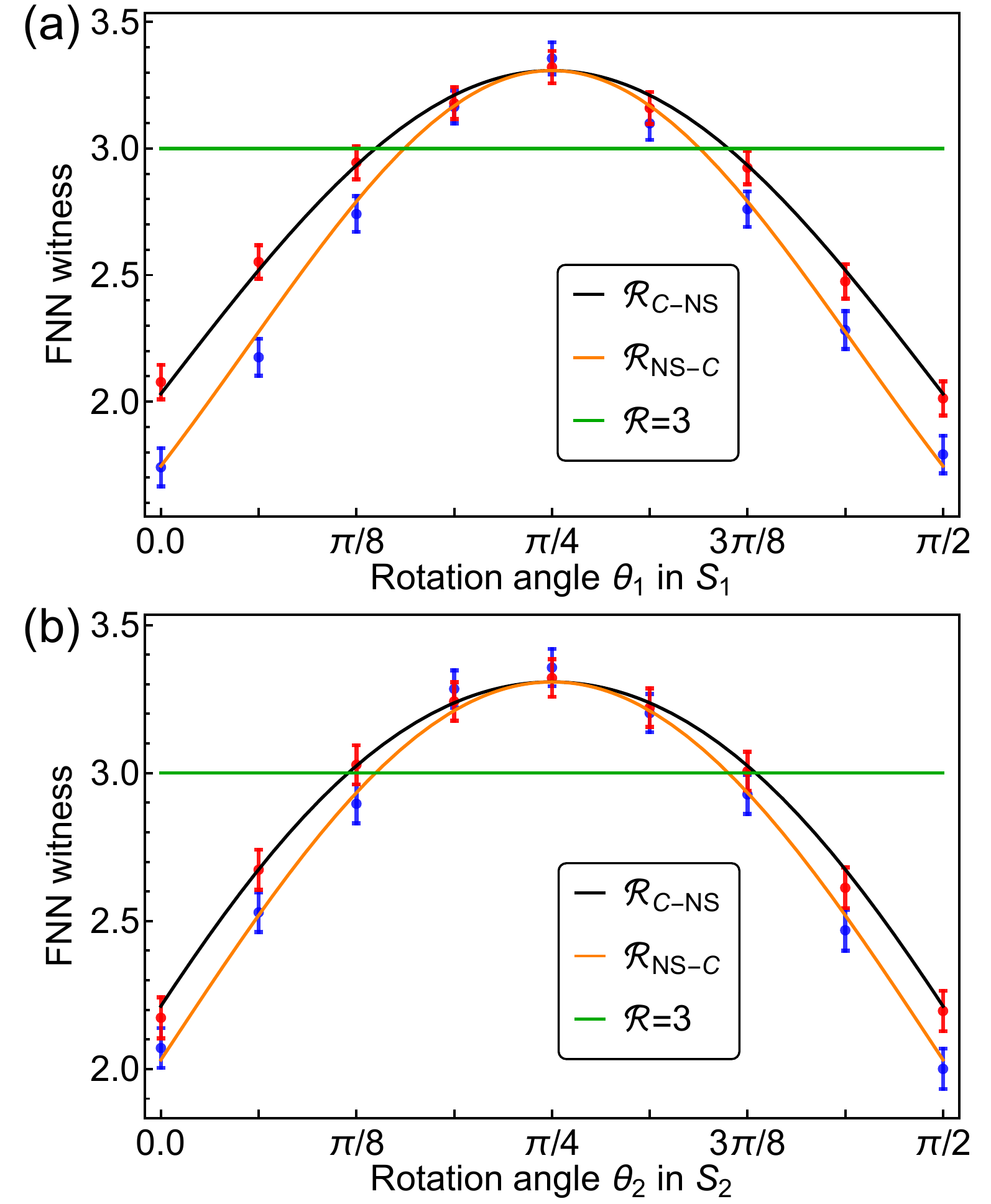}
	\caption{Experimental results of the FNN witness as functions of the rotation angle $\theta_{i}$ in $\text{S}_{i}$. \textbf{(a)} FNN witness as a function of the angle $\theta_{1}$ when $\ket{\varphi}_{2}=\ket{\Phi^{+}}$. \textbf{(b)} FNN witness as a function of the angle $\theta_{2}$ when $\ket{\varphi}_{1}=\ket{\Phi^{+}}$. Error bars indicate one standard deviation.} 
	\label{Fig:results}
\end{figure}

\paragraph*{Results.} We obtain visibilities of ($97.10 \pm 0.35$)\% and ($98.60 \pm 0.07$)\% for states $\ket{\Phi^{+}}$ generated in the diagonal polarization basis for sources $\text{S}_{1}$ and $\text{S}_{2}$, respectively, at an average photon pair number per pulse of $\sim$0.007. The Hong-Ou-Mandel measurement by Bob yields a fitted visibility of ($94.3 \pm 2.7$)\%. We then study the distributions $p(a,b,c|x,z)$ generated in our network with different states computing their corresponding values of $\mathcal{R}_{\text{C-NS}}$ and $\mathcal{R}_{\text{NS-C}}$ (see Appendix \ref{app:rotation}). For simplicity, we fix one source creating $\ket{\Phi^{+}}$ and then automatically rotate the phase $\theta_{i}$ in the other source with a motorized rotator at a phase interval of $\pi/16$, having a total of 9 steps within the range $[0,\pi/2]$. For each step, we collect $\sim$4700 four-photon coincidence detection events in $\sim$\mbox{10 000} seconds, and compute with them the results shown in Fig. \ref{Fig:results}. Our observation of $\mathcal{R}_{\text{C-NS}}>3$ in Fig.~\ref{Fig:results}(a) certifies that $\text{S}_1$ distributes an entangled quantum system to Alice and Bob. An analogous argument applies for $\text{S}_2$ and Eq.~\eqref{eq:functionF2} in Fig.~\ref{Fig:results}(b). But more importantly, the fact that in Fig. \ref{Fig:results}(a) we observe a simultaneous violation of both inequalities for the points $\theta_1=3\pi/16$, $\pi/4$ and $5\pi/16$ is, at least in those cases, a guarantee that both sources in the network are non-classical. The same conclusions can be obtained when changing the state in $\text{S}_2$ as we show in Fig. \ref{Fig:results}(b). Notably, in the particular case of $\ket{\varphi}_{1}=\ket{\varphi}_{2}=\ket{\Phi^{+}}$, our results yield $\mathcal{R}_{\text{C-NS}}=3.3212 \pm 0.0638$ and $\mathcal{R}_{\text{NS-C}}=3.3563 \pm 0.0632$ both in Fig. \ref{Fig:results} (a) and (b), surpassing the non-FNN's bound by more than five standard deviations.

\paragraph*{Discussion.} We have experimentally demonstrated the existence of FNN in a photonic quantum network built upon sharing independent sources under strict locality constraints. FNN is witnessed by the simultaneous violation of Eqs.~\eqref{eq:functionF1} and \eqref{eq:functionF2}. Each violation of certifies, in a device-independent manner, that one of the sources in the network is not classical. Thus, our experiment constitutes a certification of the non-classicality of all the sources in the entanglement-swapping scenario. Importantly, the fact that non-FNN models consider general no-signaling systems implies that, even in the hypothetical case that new physical systems were discovered that allowed for stronger-than-quantum correlations, it would remain impossible to reproduce the results of our experiments by using these systems in one source and classical systems in the other. This is a very important feature for quantum communication networks.
In contrast with pioneering work \cite{lee2018towards} which requires all the sources to distribute stronger-than-quantum systems in order to limit the information accessible by an eavesdropper, FNN could provide a strong, yet achievable in quantum theory, way of guaranteeing security of network-based quantum cryptographic protocols.

Our experiment addresses the source-independence, locality, and measurement-independence loopholes, thus providing a strong certification of FNN. However, it remains subject to the detection loophole, namely that a local model could be given if taking into account non-detection events \cite{pearle1970}, and the memory loophole, by which the results in a given experimental round may depend on the previous ones \cite{accardi,barrett2002}. These loopholes need to be closed in order for the certification to be considered device independent \cite{hensen2015loophole}. The former could be addressed in the future by using high-efficiency photon sources \cite{wang2019towards} and detectors, while closing the latter requires suitable hypothesis testing \cite{barrett2002,gill2001} and sufficiently many experimental rounds, thus benefiting of higher-frequency hardware. Beyond the bilocal scenarios, an important direction is to explore FNN in more complex networks such as star scenarios \cite{tavakoli2014nonlocal, andreoli2017maximal, tavakoli2017correlations} where several independent branch parties are connected to a central one, and line scenarios that underlie long-distance quantum communication networks.

\paragraph*{Acknowledgments.}  We thank Chang Liu and Quantum Ctek for providing the components used in the quantum random number generators. This work was supported by the National Key Research and Development Program of China (2020YFA0309704), Shandong Provincial Natural Science Foundation (ZR2021LLZ005), the National Natural Science Foundation of China (12004138), the Chinese Academy of Sciences, the National Fundamental Research Program, the Anhui Initiative in Quantum Information Technologies, the Spanish Ministry of Science and Innovation MCIN/AEI/10.13039/501100011033 and the European Union NextGenerationEU/PRTR (CEX2019-000904-S and PID2020-113523GB-I00), the Spanish Ministry of Economic Affairs and Digital Transformation (project QUANTUM ENIA), Comunidad de Madrid (QUITEMAD-CM P2018/TCS-4342), Universidad Complutense de Madrid (FEI-EU-22-06), and the CSIC Quantum Technologies Platform PTI-001.

\appendix
\section{An analytic derivation for FNN in the entanglement-swapping scenarios}\label{app:analytic}
\textbf{Theorem.} Consider the network with three measurement nodes in which Alice and Bob (or Charlie and Bob) share a classical source while Bob and Charlie (or Alice and Bob) share a non-signaling nonlocal source with Alice and Charlie measuring two dichotomic observables $ A_x$ and $ C_z$ ($x,z=0,1$) while Bob measures a single observable with three outcomes, resulting in distributions $p(a,b,c|x,z)$  with outcomes $a,c=0,1$ and $b=0,1,2$. Then it holds $\mathcal{R}_{\text{C-NS}}\leq 3 $ (or $\mathcal{R}_{\text{NS-C}}\leq 3 $) where
\begin{equation}	 
\begin{aligned} 
	\mathcal{R}_{\text{C-NS}}:= & 2\left\langle A_{0} B_{1} \left( C_{0} - C_{1} \right) \right\rangle + \left\langle A_{1} B_{0} \left( 2C_{0} + C_{1} \right) \right\rangle\\
	& - \left\langle B_{0} \right\rangle + \left(\left\langle A_{1} B_{0} \right\rangle + \left\langle B_{0} C_{0} \right\rangle - \left\langle C_{0} \right\rangle \right) \left\langle C_{1} \right\rangle, \\
	\\
	\mathcal{R}_{\text{NS-C}}:= & 2\left\langle A_{0} B_{1} \left( C_{0} - C_{1} \right) \right\rangle + \left\langle A_{1} B_{0} \left( C_{0} + 2 C_{1} \right) \right\rangle  \\
	& - \left\langle B_{0} \right\rangle + \left\langle A_{1} \right\rangle (\left\langle A_{1} B_{0} \right\rangle + \left\langle B_{0} C_{1} \right\rangle \\
	& + \left\langle C_{0} - C_{1} - A_{1} \right\rangle),
\end{aligned}
\label{eq:functionCorrelation}
\end{equation}
with $B_0 = \Pi_0 + \Pi_1 - \Pi_2 $ and $ B_1 = \Pi_0 - \Pi_1$ \cite{branciard2012bilocal}, e.g., 
\begin{equation*}
 \begin{aligned}
 	 &\mean{A_x B_0 C_z} =\sum_{a,c}(-1)^{a+c}[p(a,0,c|x,z)+p(a,1,c|x,z)\\
 	&\hspace{2.cm}-p(a,2,c|x,z)],\\
 	&\mean{A_x B_1 C_z}\! = \!\!\sum_{a,c} (-1)^{a+c}\left[p(a,0,c|x,z)\!-\!p(a,1,c|x,z)\right].
 \end{aligned}
\label{eq:Threeparty}
\end{equation*} 

\begin{figure}[!t]
	\centering
	\includegraphics[width=0.46\textwidth]{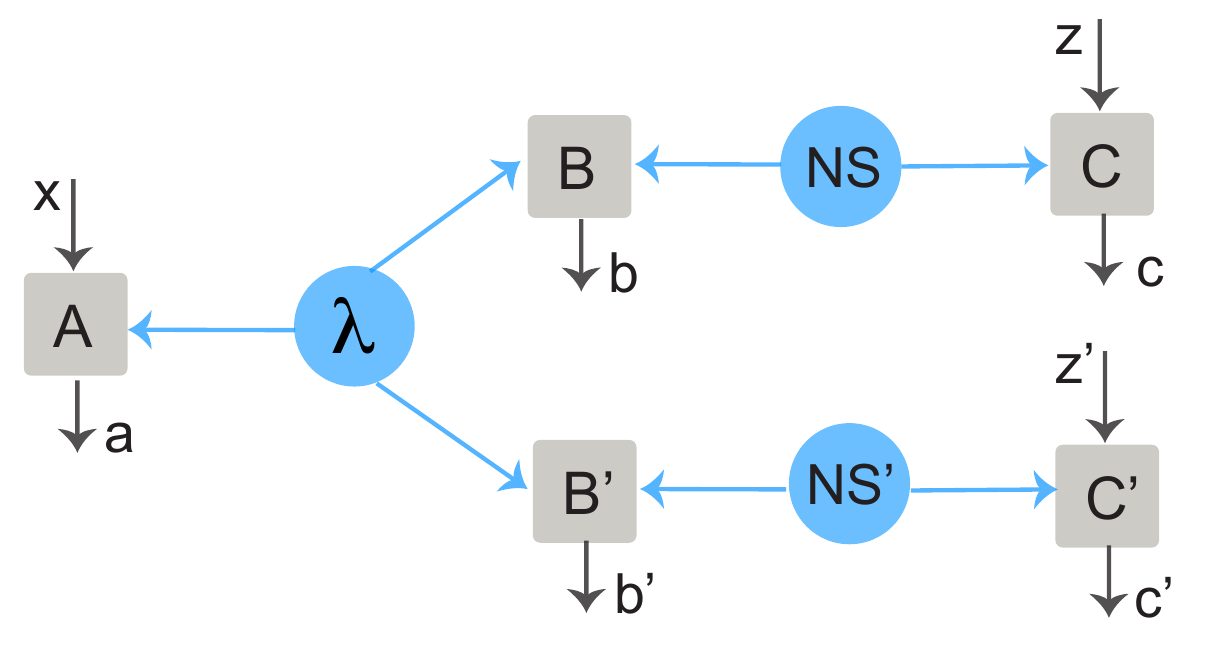}
	\caption{Inflated network of the entanglement-swapping scenario in Fig.1 (b) of the main text. The source denoted by $\lambda$ distributes a classical system to Alice and Bob. This classical information can be cloned and sent to copies of Bob (in the figure, B and B'). The subsystems of the general nonlocal resource between Bob and Charlie cannot be cloned due to the no-cloning theorem, but one can consider duplicate copies that connect different copies of Bob and Charlie (e.g., by going to the manufacturer and buying a new source and measurement devices) \cite{wolfe2019inflation, gisin2020constraints}.} 
	\label{Fig:inflatenetwork}
\end{figure}

\textbf{Proof.} The proof will be based on inflation arguments~\cite{wolfe2019inflation,wolfe2021qinflation}. Namely, in order to constrain $\mathcal{R}_\text{C-NS}$ under distributions of the type of Eq. (1) in the main text, we will consider more complex scenarios built from their elements. For instance, in the inflated network of Fig. \ref{Fig:inflatenetwork} we have correlations $Q(abb'cc'|xzz')$ satisfying a set of compatibility constraints, some of which read
\begin{equation*}
	\begin{aligned}
		&\mean{A_x B_y C'_z} _Q = \mean{A_x B_y} \mean{C_z},\\
		&\mean{A_x C'_z} _Q = \mean{A_x C_z} _Q = \mean{A_x C_z}.
	\end{aligned}	
\end{equation*}

As a result we have 
\begin{equation*}
	\begin{aligned}
		&\left( \mean{A_1 B_0} + \mean{B_0 C_0} - \mean{C_0} \right) \mean{C_1}\\
		&= \mean{A_1 B_0 C'_1 + B_0 C_0 C'_1 - C_0 C'_1} _Q.
	\end{aligned}	
\end{equation*}	

Let $p_b$ be the probability of obtaining results $ b = 0, 1, 2 $ by Bob measuring his systems and denote by $ \mean{\cdot}_b $ the corresponding conditional average for subsystems Alice and Charlie. Then according to the outcome of Bob's measurement we have the decomposition
\begin{equation*}
	\mathcal{R}_{\text{C-NS}}= \sum_{j=0}^{2} p_b T_b
\end{equation*} 
with
\begin{equation*}
	\begin{aligned}
		T_b \coloneqq &\, \mean{2 A_0 (C_0 - C_1)(-1)^b + A_1 (2C_0 + C_1 + C'_1)}_b -\!1\\
		=&\, 2\mean{A_0 (C_0 - C_1)(-1)^b + A_1(C_0 + C_1)}_b -1 \\
		&\hspace{0.5cm}+\mean{A_1 (C'_1 - C_1)}_b\\
		\leq&\, 3 + \mean{A_1 (C'_1 - C_1)}_b
	\end{aligned}
\end{equation*}
for $ b = 0,1$, and 
\begin{equation*}
	\begin{aligned}
		T_2 \coloneqq &\, 1 - \mean{A_1 (2 C_0 + C_1 +C'_1) + 2C_0 C'_1} _2\\
		=&\, 1 + \mean{A_1 (C'_1 - C_1)} _2 - 2 \mean{A_1 (C_0 + C'_1) + C_0 C'_1} _2\\
		\leq&\, 3 + \mean{A_1 (C'_1 - C_1)} _2,
	\end{aligned}
\end{equation*}
where we for the first we have used the fact that the variable $\lambda$ is classical and thus $\mean{A_0 (C_0 - C_1)(-1)^b + A_1(C_0 + C_1)}_b\leq 2$, and for the second we have used the inequality $ 1 + \mean{A_1 (C_0 +C'_1) + C_0 C'_1} \geq 0 $ which comes from positivity:
\begin{equation*}
	\begin{aligned}
		2^3 &\sum_{\alpha=0,1} q(a_1=\alpha,c_0=\alpha,c'_1=\alpha)  \\
		&=\!\!\!\!\sum_{\alpha=0,1}\!\! \mean{(\mathbb{1}\! +\! (-1) ^ \alpha A_1) (\mathbb{1}\! +\! (-1) ^ \alpha C_0) (\mathbb{1}\!+\! (-1) ^ \alpha C'_1)} _Q \\
		&= 2\mean{\mathbb{1} + A_1(C_0+C'_1) + C_0C'_1}_P\geq 0.
	\end{aligned}
\end{equation*}

By noting that
\begin{equation*}
	\sum_{b=0}^{2} p_b \mean{A_1 (C'_1 - C_1)}_b = \mean{A_1 (C'_1 - C_1)} _Q = 0,
\end{equation*}
which is due to inflation \cite{wolfe2019inflation, gisin2020constraints}, we obtain $ \mathcal{R}_{\text{C-NS}} \leq 3$.

Similarly, we have the decomposition 
\begin{equation*}
	\mathcal{R}_{\text{NS-C}}=\sum_{b=0}^{2} p_b S_b
\end{equation*} with 
\begin{equation*}
	\begin{aligned}
	S_b&\coloneqq\!\mean{2 A_0 (C_0\!-\!C_1)(-1)^b\!+\!A_1(C_0\!+\!2 C_1)\!+\!A'_1 C_0}_b \!-\!1\\
     &=2 \mean{A_0 (C_0 - C_1) (-1)^b +A_1(C_0 +C_1)}_b - 1 \\
     &\hspace{0.5cm}+\mean{(A'_1 - A_1) C_0}_b\\
     &\leq 3 + \mean{(A'_1 - A_1) C_0}_b
	\end{aligned}
\end{equation*}
for $ b=0,1$, and
\begin{equation*}
	\begin{aligned}
	S_2&\coloneqq 1 - \mean{A_1 (C_0 +2 C_1) + 2 A'_1 (A_1 + C_1) - A'_1 C_0}_2\\
	&= 1 + \mean{(A'_1 - A_1) C_0} - 2 \mean{C_1 (A'_1 + A_1) + A_1 A'_1}_2\\
	&\leq 3 + \mean{(A'_1 - A_1) C_0}_2,
	\end{aligned}
\end{equation*}
leading to $ \mathcal{R}_{\text{NS-C}}\leq 3 + \sum_{b} p_b \mean{(A'_1 - A_1) C_0}_b = 3$.

\section{The FNN witnesses as functions of rotation angle $\theta_{i}$ in the source $\text{S}_{i}$}\label{app:rotation}
As described in the main text, sources $\text{S}_{1}$ and $\text{S}_{2}$ distribute entangled polarization states $\ket{\varphi_{1}}$ and $\ket{\varphi_{2}}$ to Alice and Bob, and to Bob and Charlie, respectively. We model the state $\ket{\varphi_{i}}$ produced by the source $\text{S}_{i}$ with a visibility  $v_{i}$ as the following:
\begin{equation}
	\rho_{i}=v_{i}\ket{\varphi_{i}}\bra{\varphi_{i}}+\frac{1-v_{i}}{4}\mathbb{1},
\label{eq:states}
\end{equation}
where $\ket{\varphi_{i}}=\cos{2\theta_{i}}\ket{HH}+\sin{2\theta_{i}}\ket{VV}$, $\ket{H}$ and $\ket{V}$ respectively represent the horizontal and vertical polarization states, and $\theta_{i}$ is the angle of the half-wave plate $\text{HWP}_{i}$ in source $\text{S}_{i}$, as shown in Fig.2(b) of the main text. When the rotation angles are $\theta_{1}=\theta_{2}=\pi/4$ the visibilities are $v_{1}=v_{2}=1$, the created states are the Bell state $\ket{\Phi^{+}}$. 

To witness full network nonlocality (FNN) in the experiment, Alice and Charlie perform measurements $A_{x}\in\{X,Z\}$ and $C_{z}\in\{\frac{Z+X}{\sqrt{2}},\frac{Z-X}{\sqrt{2}}\}$ with binary inputs $x,z\in\{0,1\}$ yielding outcomes $a,c=0,1$, respectively. Bob performs a partial Bell state measurement (BSM) yielding the three outcomes $\{\Phi^{+}, \Phi^-, \mathbb{1}-\Phi^{+}-\Phi^-\}$ that are correspondingly associated to the outcomes $b\in\{0,1,2\}$, where $\Phi^{\pm}=\ket{\Phi^{\pm}}\bra{\Phi^{\pm}}$. The partial BSM by Bob can be modeled as the following 
\begin{equation}
\begin{aligned}
&\Pi_{0}=v_{h}\ket{\Phi^+}\bra{\Phi^+}\!+\!\frac{(1-v_{h})}{2}(\ket{\Phi^+}\bra{\Phi^+}+\ket{\Phi^-}\bra{\Phi^-}),\\
&\Pi_{1}=v_{h}\ket{\Phi^-}\bra{\Phi^-}\!+\!\frac{(1-v_{h})}{2}(\ket{\Phi^+}\bra{\Phi^+}+\ket{\Phi^-}\bra{\Phi^-}),\\
&\Pi_{2}=\mathbb{1}-\Pi_{0}-\Pi_{1},
\end{aligned}
\label{eq:BSM}
\end{equation}
where $v_{h}$ is the visibility of the Hong-Ou-Mandel interference with two photons from sources $\text{S}_{1}$ and $\text{S}_{2}$. 

The joint input-output distributions $p(a,b,c|x,z)$ observed by the three parties in the entanglement-swapping network are 
\begin{equation}
\begin{aligned}
&p(a,b,c|x,z)\\
&=\!\text{tr}\!\left[\! \left(\frac{\mathbb{1}+(-1)^aA_{x}}{2}\otimes \Pi_{b}\otimes \frac{\mathbb{1}+(-1)^cC_{z}}{2}\right)\!\left(\rho_{1}\!\otimes\!\rho_{2}\right)\right]\!,
\end{aligned}
\label{eq:distributions}
\end{equation}
which leads to $\mathcal{R}_{\text{C-NS}}$ and $\mathcal{R}_{\text{NS-C}}$ as described with expectation values in Eq. \ref{eq:functionCorrelation}. The rest one-, two-party expectation values in Eq. \ref{eq:functionCorrelation} is the following \cite{pozas2022full}:
\begin{equation*}
    \begin{aligned}
     & \mean{A_x} = \sum_{a,b,c} (-1)^{a}p(a,b,c|x,z), \\
     & \mean{C_z} =\sum _{a,b,c} (-1)^{c}p(a,b,c|x,z),\\
       & \mean{B_1} = \sum_{a,c} \left[p(a,0,c|x,z)-p(a,1,c|x,z)\right],\\ 
       &  \mean{B_0} = \sum_{a,c} \left[p(a,0,c|x,z)+p(a,1,c|x,z)\right. \\
       &\qquad\qquad\quad\left.-p(a,2,c|x,z)\right],\\
        & \mean{A_x C_z} = \sum_{a,b,c} (-1)^{a+c}[p(a,b,c|x,z)],\\
		&  \mean{A_x B_0} = \sum_{a,c} (-1)^{a}\left[p(a,b,c|x,z)+p(a,1,c|x,z)\right. \\
		&\qquad\qquad\qquad\qquad\quad\left.-p(a,2,c|x,z)\right],\\
		& \mean{B_0 C_z} =\sum_{a,c} (-1)^{c}\left[p(a,0,c|x,z)+p(a,1,c|x,z)\right. \\
		&\qquad\qquad\qquad\qquad\quad\left.-p(a,2,c|x,z)\right],\\
		 &   \mean{A_x B_1} = \sum_{a,c} (-1)^{a}\left[p(a,b,c|x,z)-p(a,1,c|x,z)\right],\\
       &  \mean{B_1 C_z} = \sum_{a,c} (-1)^{c}\left[p(a,0,c|x,z)-p(a,1,c|x,z)\right].
	    \end{aligned}
   	\label{eq:expectionValues}
\end{equation*}

By substituting Eqs. \ref{eq:states}, \ref{eq:BSM} and \ref{eq:distributions} into Eq. \ref{eq:functionCorrelation} with the above described expectation values, we have
\begin{equation}
	\begin{aligned}
		\mathcal{R}_{\text{C-NS}}:=&\frac{3v_{1}v_{2}}{\sqrt{2}}-\frac{1}{2}v_{2}^2\cos^2{4\theta_{2}}+\frac{v_{1}v_{2}^2\cos^2{4\theta_{2}}}{\sqrt{2}}\\
		&-v_{1}v_{2}\cos{4\theta_{1}}\cos{4\theta_{2}}\!+\!\frac{1}{2}v_{1}v_{2}^2\cos{2v_{1}}\cos{2v_{2}}\\
		&+\sqrt{2}v_{1}v_{2}v_{\text{h}}\sin{4\theta_{1}}\sin{4\theta_{2}}
	\end{aligned}
	\label{eq:functionFS1}
\end{equation}
and 
\begin{equation}
	\begin{aligned}
		\mathcal{R}_{\text{NS-C}}:=&\frac{3v_{1}v_{2}}{\sqrt{2}}-v_{1}^2\cos^2{4\theta_{1}}+\frac{v_{1}^2v_{2}\cos^2{4\theta_{1}}}{\sqrt{2}}\\
		&-v_{1}v_{2}\cos{4\theta_{1}}\cos{4\theta_{2}}+v_{1}^2v_{2}\cos{4\theta_{1}}\cos{4\theta_{2}}\\
		&+\sqrt{2}v_{1}v_{2}v_{\text{h}}\sin{4\theta_{1}}\sin{4\theta_{2}}.
	\end{aligned}
	\label{eq:functionFS2}
\end{equation}

\begin{figure}[h]
	\centering
	\includegraphics[width=0.48\textwidth]{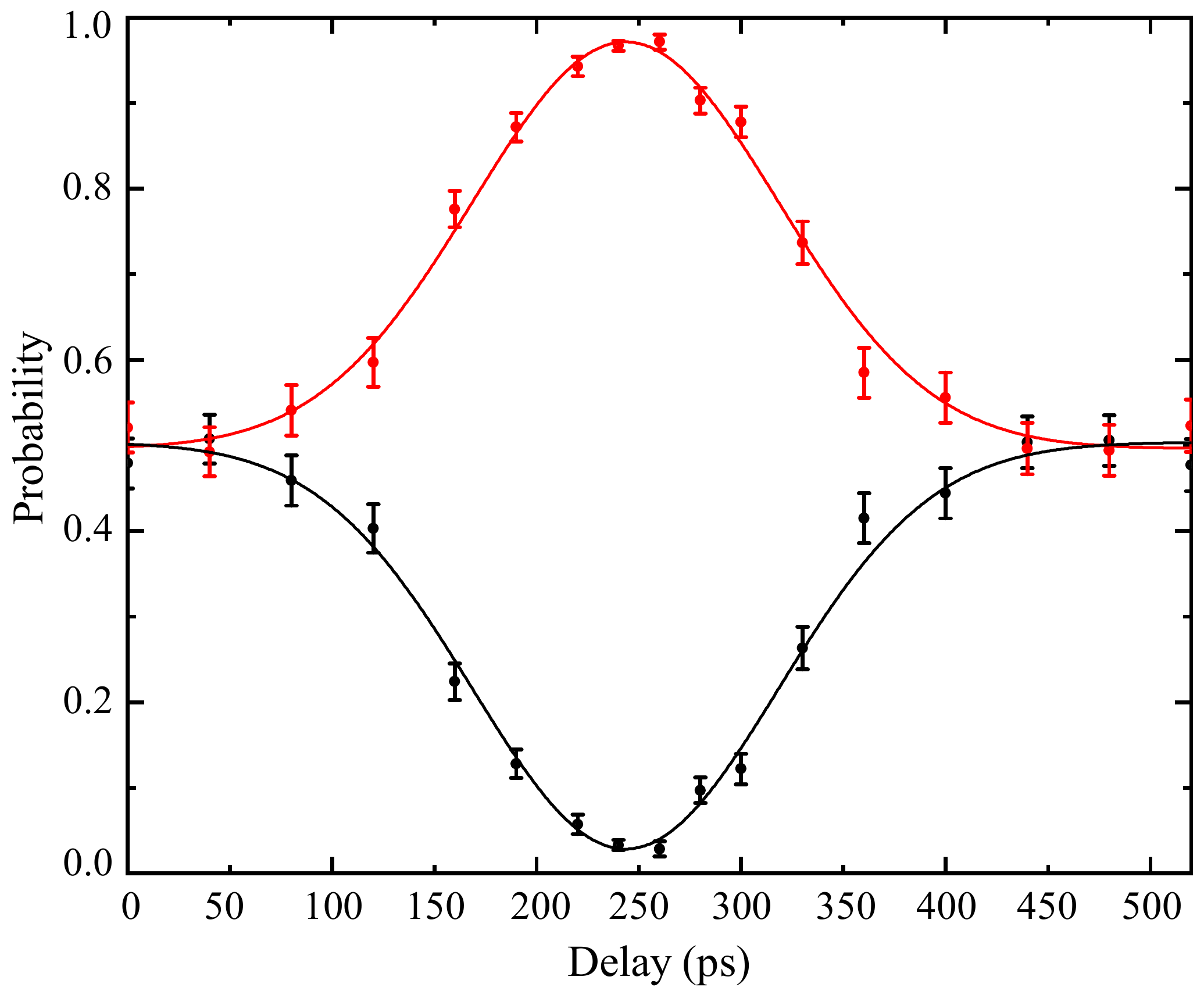}
	\caption{Experimental result of the Hong-Ou-Mandel interference. Each data point is accumulated for more than 900s and the error bar represent a standard deviation in the experiment. The fitted visibility is ($94.3 \pm 2.7$)\%.} 
	\label{Fig:HOM}
\end{figure}

We set $\theta_{i}=\pi/4$ and prepared both states in the Bell state $\ket{\varphi}_{1}=\ket{\varphi}_{2}=\ket{\Phi^{+}}$. The obtained visibility of the Bell state in diagonal polarization bases for source $\text{S}_{1}$ and $\text{S}_{2}$ are $v_{1}=(97.10 \pm 0.35)$\% and $v_{2}=(98.60 \pm 0.07)$\%, respectively. We perform a Hong-Ou-Mandel measurement at Bob's station on the photons from the two separate sources $\text{S}_{1}$ and $\text{S}_{2}$. By suppressing the distinguishability of photons in spectral, spatial, temporal and polarization modes, we obtained a fitted visibility of $v_{h}=(94.3 \pm 2.7)$\%, as shown in Fig. \ref{Fig:HOM}.

\section{Measurements by Alice, Bob and Charlie}\label{app:qrng}
At Alice's and Charlie's station, a high-speed polarization analyzer is implemented for measuring their received photons from the sources, which consists of two fixed quarter-wave plates (QWPs$@45^{\circ}$), two Faraday rotators (FRs) and an electro-optical phase modulator (PM), as described in Fig.2(c) of the main text. When a single photon impinges onto the loop interferometer, two of its orthogonal polarization components exit at different ports of the PBS. We use the FR to rotate the single photon's polarization and align it with the slow axis of a polarization-maintaining fiber connected to the PM, such that the two polarization components are coupled into the fiber in the loop. The PM is displaced from the loop center by 20 cm to create a relative delay of $\sim 2$ ns between the arrival times of the two counter-propagating components at the PM, allowing us to adjust the phase of a single component individually. In the end, the two components interfere at the PBS and exit as a single photon pulse with a modulated polarization state. The phase of the PM is controlled by a quantum random number generator (QRNG) situated there, allowing to switch between measurement settings $A_{x}$ for Alice and $C_{z}$ for Charlie, respectively. The corresponding configurations are shown in Table \ref{tab:settings}.
\begin{table}[h]
	\caption{Alice and Charlie perform measurement $A_{x}$ and $C_{z}$ according to the random inputs $x,z\in\{0,1\}$ from their private QRNG.}
	\renewcommand{\arraystretch}{1.5}
	\begin{tabular}{|>{\centering}m{1.9em}|c|c|>{\centering}m{3em}|c|}
		\hline
		\multicolumn{1}{|c|}{$x$} & QWP   & $\phi_p$ & QWP  & $A_x$  \\ \hline
		0   & $@-45^{\circ}$ & $@ 90^{\circ}$  & $@45^{\circ}$ & $X$    \\ \hline
		1   & $@-45^{\circ}$ & $@ 0^{\circ}$  & $@45^{\circ}$ & $Z$   \\ \hline
		\hline
		\multicolumn{1}{|c|}{$z$} & QWP  & $\phi_p$ & QWP  & $C_z$   \\ \hline
		0   & $@-45^{\circ}$ & $@ 45^{\circ}$ & $@45^{\circ}$ & $(Z+X)/\sqrt{2}$   \\ \hline
		1   & $@-45^{\circ}$ & $@-45^{\circ}$ & $@45^{\circ}$ & $(Z-X)/\sqrt{2}$    \\ \hline
	\end{tabular}
	\label{tab:settings}
\end{table}

At Bob's station, a partial BSM is implemented to perform projection measurement $\Pi_{0}$,  $\Pi_{1}$, and $\Pi_{2}$, yielding three outcomes $\Phi^+$, $\Phi^-$, and $\mathbb{1}-\Phi^+-\Phi^-$ respectively. Two superconducting nanowire single photon detectors (SNSPDs), and a 50:50 beam splitter (BS) is used as a pseudo-number-resolving detectors (denoted as $\text{D}_{\text{H}1}$, $\text{D}_{\text{V}1}$, $\text{D}_{\text{H}2}$, and $\text{D}_{\text{V}2}$). The three outcomes are discriminated by the corresponding two-photon coincidence detection between these SNSPDs, as shown in Table \ref{tab:Bobphotonclick}.
\begin{table}[h]
	\caption{Coincidence detection events between the detectors at Bob's location, lead to three outcomes.}
	\renewcommand{\arraystretch}{1.5}
	\begin{tabular}{|c|c|c|}
		\hline
		\multicolumn{1}{|c|}{Projector} & Coincidence detection events& Outcomes   \\ \hline
       $\Pi_{0}$& ($\text{D}_{\text{H}1}$ \& $\text{D}_{\text{H}2}$) or ($\text{D}_{\text{V}1}$ \& $\text{D}_{\text{V}2})$   &  $\Phi^+$  \\ \hline
		$\Pi_{1}$& $(\text{D}_{\text{H}1}$ \& $\text{D}_{\text{V}2}$) or ($\text{D}_{\text{V}1}$ \& $\text{D}_{\text{H}2}$)  &  $\Phi^-$ \\ \hline
		$\Pi_{2}$&$\text{D}_{\text{H}1}$ or $\text{D}_{\text{V}2}$ or $\text{D}_{\text{V}1}$ or $\text{D}_{\text{H}2}$  &  $\mathbb{1}-\Phi^{+}-\Phi^{-}$ \\ \hline
		\multicolumn{3}{p{8cm}}{Notes: For $\Pi_{0}$ and $\Pi_{1}$, detection of $D_{\text{H}_{1}}$ means that only one of its two composed SNSPDs click. This is the same for the detection of $\text{D}_{\text{V}1}$, $\text{D}_{\text{H}2}$, and $\text{D}_{\text{V}2}$. For $\Pi_{2}$, detection of $D_{\text{H}_{1}}$ means that its two composed SNSPDs both click simultaneously. This is the same for the detection of $\text{D}_{\text{V}1}$, $\text{D}_{\text{H}2}$, and $\text{D}_{\text{V}2}$. The detection for outcome $\mathbb{1}-\Phi^{+}-\Phi^{-}$ has a success probability of 50\%.}
	\end{tabular}
	\label{tab:Bobphotonclick}
\end{table}

\section{The space-time analysis}\label{app:spacetime}
We characterize three important types of events: (1) emission, i.e., the photon creation in the source $\text{S}_{1}$ and $\text{S}_{2}$. (2) the choice of measurement setting, i.e., completing the quantum random number generation for setting choices at Alice's node ($\text{QRNG}_{A}$) and Charlie's node ($\text{QRNG}_{C}$). (3) measurement, i.e., finishing the single photon detection by Alice ($\text{M}_{A}$), Bob ($\text{M}_{B}$), and Charlie ($\text{M}_{C}$). To meet the strict locality constraints, we space-like separate all the relevant events from the above characterizations. That is, one event is outside light cone of other events such that no information can transmit from one to the other within the duration of the experiment. More specifically, space-like separation means that two events whose space distance $d$ and time interval $ \Delta t$ satisfy the following inequality \cite{jackson1999classical}
\begin{equation}
	\Delta s^2 = d^2 -c^2 \Delta t^2 > 0. \label{eq:space-like}
\end{equation}

We measure the space distances and fiber length for adjacent nodes  in our experiment. The size of each measurement node (on an optical table) is less than 1 m. The spatial distance between these nodes is precisely measured by a laser rangefinder with uncertainty within 0.4 cm, which is much smaller than the size of each node. This allows us to put an upper bound of the distance uncertainty to be 1 m.  The fiber links are measured with a ruler in advance, and we set the uncertainty to 0.1 m taking the fiber fusion accuracy into account. The angles shown in Fig. 2(a) are used to calculate other beeline distances between the nonadjacent nodes in the setup. All lengths are listed in Table \ref{table:lengths}.
\begin{table}[!h]
	\caption{Space distances and fiber links between the adjacent nodes in our network.}
	\renewcommand{\arraystretch}{1.5}
	\begin{tabular}{|c|c|c}
		\hline
		Link         & Space distance (m) & \multicolumn{1}{c|}{Fiber length (m)} \\ \hline
		Alice -- $\text{S}_1$      & $ L_{A1} = 110 \pm 1$        & \multicolumn{1}{c|}{$ l_{A1}=125.48 \pm 0.1$}  \\ \hline
		$\text{S}_1$ -- Bob        & $ L_{B1}=89  \pm 1$        & \multicolumn{1}{c|}{$ l_{B1}=108.75 \pm 0.1$}  \\ \hline
		Bob -- $\text{S}_2$        & $ L_{B2}=106 \pm 1$        & \multicolumn{1}{c|}{$ l_{B2}=123.86 \pm 0.1$}  \\ \hline
		$\text{S}_2$ -- Charlie    & $ L_{C2}=104 \pm 1$        & \multicolumn{1}{c|}{$ l_{C2}=112.63 \pm 0.1$}  \\ \hline 
		\multicolumn{3}{p{8cm}}{Space distances between the nonadjacent nodes.}\\ \cline{1-2}
		Alice -- Bob     & $ L_{AB}=199 \pm 1$   &        \\ \cline{1-2}
		Bob -- Charlie   & $ L_{BC}=192 \pm 1$   &          \\ \cline{1-2}
		Alice --Charlie & $ L_{AC}=384 \pm 1$    &          \\ \cline{1-2}
		Alice -- $\text{S}_2$     & $ L_{A2}=306 \pm 1$        &          \\ \cline{1-2}
		$\text{S}_1$ -- $\text{S}_2$         & $ L_{12}=195 \pm 1$    &      \\ \cline{1-2}
		$\text{S}_1$ -- Charlie    & $ L_{C1}=277 \pm 1$    &     \\ \cline{1-2}
	\end{tabular}
	\label{table:lengths}
\end{table}

We now define each experimental trial with start time and end time marked by the pump pulse generated in source $\text{S}_{2}$ and a signal output from the SNSPDs, respectively. To have the occurrence time of all events, we measure the related indispensable delays, as shown in Table \ref{table:otherdealys}. Here, $\Delta T_{\text{S}_{i}} $ denotes the time interval between the source $\text{S}_{i}$ generating a pump laser pulse and entangled photons produced from the source entering into the fiber coupler, $ \Delta T_{\text{QRNG}_{i}}$ is the time lapse for a quantum random number generator to produce and deliver a setting choice by generating random bits, $\Delta T_{\text{QM}_i} $ and $ \Delta T_{\text{M}_i} $ refer to the electrical delay before the random bit arrives at the PM and the optical delay from the PM to SNSPDs at node $i$, receptively.
\begin{table}[!h]
	\caption{The delay for related events in the network}
	\renewcommand{\arraystretch}{1.5}	
	\begin{tabular}{|m{5.em}|c|}
		\hline
		\multicolumn{2}{|c|}{Delays (ns)}  \\  \hline
		$ \Delta T_{\text{S}_1} $ & $154.4\pm0.5$ \\ \hline
	    $ \Delta T_{\text{S}_2} $ & $160\pm0.5$  \\ \hline
		$ \Delta T_{\text{QRNG}_{A}}$ & $53\pm2 $     \\ \hline
	   $ \Delta T_{\text{QRNG}_{C}} $ & $53\pm2$      \\ \hline
		$ \Delta T_{\text{QM}_A} $  & $446.8\pm0.5$ \\ \hline
		$ \Delta T_{\text{QM}_C} $ & $408.6\pm0.5$ \\ \hline
		$ \Delta T_{\text{M}_A} $  & $38.4\pm0.5$  \\ \hline
	   $ \Delta T_{\text{M}_B} $  & $44.9\pm0.5$ \\ \hline
		$ \Delta T_{\text{M}_C} $   & $44.6\pm0.5$\\ \hline
	\end{tabular}\label{table:otherdealys}
\end{table}

\begin{figure}[!h]
	\centering
	\includegraphics[width=0.5\textwidth]{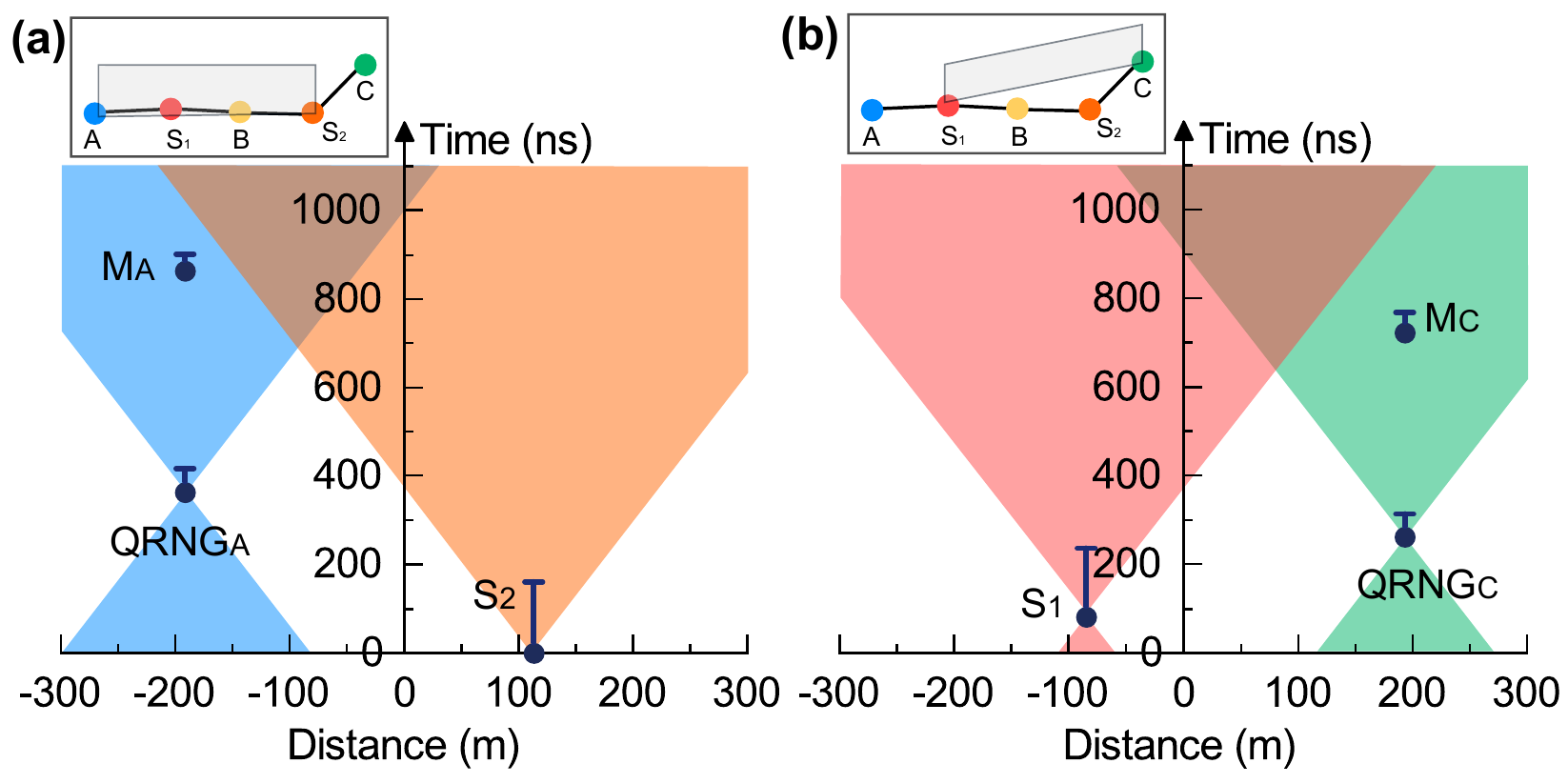}
	\caption{Space-like separations between the measurements and emission events at the sources. (a), Space-like separation between Alice's measurement $\text{M}_{\text{A}}$ and emission event at source $\text{S}_{2}$. (b), Space-like separation between Charlie's measurement $\text{M}_{\text{C}}$ and emission event at source $\text{S}_{1}$.}
	\label{Fig:SpacetimeMS}
\end{figure}

Photons from the independent source arrive at Bob's measurement station simultaneously (for good interference), allowing us to determine the time of individual measurement events in the network calculated from the fiber links in Table \ref{table:lengths} and optical delays in Table \ref{table:otherdealys}. Additionally, as the random bit from the QRNG and the photon arrive at the PM simultaneously, we can also derive the moment of choosing the measurement setting for Alice and Charlie, respectively. For example, Alice's measurement time can be calculated as $T_{\text{M}_A}=\Delta T_{\text{S}_2} + (l_{\text{B2}} - l_{\text{B1}} + l_{\text{A1}}) / v_{f} = 862.95 \pm 1~$(ns), where $v_{f} = 0.2~$(m/ns) is the speed of light in optical fibers. Similarly, the earliest time (start time) for Alice chose a measurement setting based on the random numbers can be calculated as $ T_{\text{sQRNG}_A} =T_{\text{M}_A}-\Delta T_{\text{QM}_A} - \Delta T_{\text{QG}_A} = 363.15 \pm 2.3~$(ns) and the moment for Alice finishing measurement is $ T_{\text{eM}_A} = T_{\text{M}_A}+\Delta T_{\text{M}_A} = 901.35 \pm 1.1 ~$(ns). In this way, we can derive all starting and ending times of the events and plot the space-like separation between Alice's measurement (Charlie's measurement) and emission at source $\text{S}_2$ ($\text{S}_1$) in Fig. \ref{Fig:SpacetimeMS}, which is also shown in Fig. 3(b) of the main text. We take the maximum time interval between the relevant events for $ \Delta t$, which is further used to infer the space-separation criterion in Eq. \ref{eq:space-like} with the speed of light $ c =0.299792 ~$m/ns. We show all the results in Table \ref{table:seperation result}, which ensures that all relevant events are well space-like separated.
\begin{table}[!h]
	\caption{Space-time separation results}
	\renewcommand{\arraystretch}{1.5}
	\begin{tabular}{|c>{\centering}m{4em}|>{\centering}m{4.5em}|c|c|}
		\hline
		\multicolumn{2}{|c|}{Events} & $d$ (m)  & $ \Delta t$ (ns) & $\Delta s^2$ \\ \hline
		\multicolumn{1}{|c|}{$\text{S}_{1}$ }& $\text{S}_{2}$ & $ 195 \pm 1$ &  $ 235.55 \pm 0.87 $ & $ 33038 \pm 392 $ \\ \hline

		\multicolumn{1}{|c|}{\multirow{5}{*}{$\text{QRNG}_{\text{A}}$}} & $\text{S}_{1}$ & $ 110 \pm 1$ & $ 335.0 \pm 0.87 $  & $ 2013 \pm 226 $   \\ \cline{2-5} 
		\multicolumn{1}{|c|}{}&$\text{S}_{2}$ & $ 306 \pm 1$ &  $ 416.15 \pm 1.12 $  & $ 78071 \pm 617 $  \\  \cline{2-5}
		\multicolumn{1}{|c|}{}&$\text{M}_{\text{B}}$  & $ 199 \pm 1$ & $ 461.05 \pm 2.24 $  & $ 20496 \pm 439 $ \\  \cline{2-5}
		\multicolumn{1}{|c|}{}&$\text{M}_{\text{C}}$ & $ 384 \pm 1$ & $ 404.6 \pm 2.35 $  & $ 132743 \pm 787 $ \\ \cline{2-5}
		\multicolumn{1}{|c|}{}& $\text{QRNG}_{\text{C}}$&  $384 \pm 1 $ & $154.6 \pm 2.35 $ & $ 145308 \pm 771$ \\ \hline

  		\multicolumn{1}{|c|}{$\text{M}_{A}$ }& $\text{S}_{2}$ & $ 306 \pm 1$ &  $ 901.35 \pm 1.12 $ & $ 20618 \pm 638 $ \\ \hline
    
		\multicolumn{1}{|c|}{\multirow{4}{*}{$\text{QRNG}_{\text{C}}$}} & $\text{S}_{1}$ & $ 277 \pm 1$ & $ 233.4 \pm 1.12 $  &   $ 71833 \pm 556 $    \\  \cline{2-5}
		\multicolumn{1}{|c|}{}& $\text{S}_{2}$  & $ 104 \pm 1$ & $ 314.55 \pm 0.87 $  &  $ 1924 \pm 214 $ \\  \cline{2-5}
		\multicolumn{1}{|c|}{}& $\text{M}_{\text{A}}$ & $ 384 \pm 1$ &  $ 639.8 \pm 2.35 $ & $ 110666 \pm 814 $ \\  \cline{2-5}
		\multicolumn{1}{|c|}{}& $\text{M}_{\text{B}}$ & $ 192 \pm 1$ & $ 562.65 \pm 2.24 $ & $ 8412 \pm 446 $ \\  \hline

  		\multicolumn{1}{|c|}{$\text{M}_{C}$ }& $\text{S}_{1}$ & $ 277 \pm 1$ &  $  686.6 \pm 1.12 $ & $ 34360 \pm 571 $ \\ \hline
	\end{tabular}\label{table:seperation result}
\end{table}

\clearpage

\bibliographystyle{apsrev4-2}
\bibliography{refs}

\begin{thebibliography}{44}%
\makeatletter
\providecommand \@ifxundefined [1]{%
 \@ifx{#1\undefined}
}%
\providecommand \@ifnum [1]{%
 \ifnum #1\expandafter \@firstoftwo
 \else \expandafter \@secondoftwo
 \fi
}%
\providecommand \@ifx [1]{%
 \ifx #1\expandafter \@firstoftwo
 \else \expandafter \@secondoftwo
 \fi
}%
\providecommand \natexlab [1]{#1}%
\providecommand \enquote  [1]{``#1''}%
\providecommand \bibnamefont  [1]{#1}%
\providecommand \bibfnamefont [1]{#1}%
\providecommand \citenamefont [1]{#1}%
\providecommand \href@noop [0]{\@secondoftwo}%
\providecommand \href [0]{\begingroup \@sanitize@url \@href}%
\providecommand \@href[1]{\@@startlink{#1}\@@href}%
\providecommand \@@href[1]{\endgroup#1\@@endlink}%
\providecommand \@sanitize@url [0]{\catcode `\\12\catcode `\$12\catcode
  `\&12\catcode `\#12\catcode `\^12\catcode `\_12\catcode `\%12\relax}%
\providecommand \@@startlink[1]{}%
\providecommand \@@endlink[0]{}%
\providecommand \url  [0]{\begingroup\@sanitize@url \@url }%
\providecommand \@url [1]{\endgroup\@href {#1}{\urlprefix }}%
\providecommand \urlprefix  [0]{URL }%
\providecommand \Eprint [0]{\href }%
\providecommand \doibase [0]{https://doi.org/}%
\providecommand \selectlanguage [0]{\@gobble}%
\providecommand \bibinfo  [0]{\@secondoftwo}%
\providecommand \bibfield  [0]{\@secondoftwo}%
\providecommand \translation [1]{[#1]}%
\providecommand \BibitemOpen [0]{}%
\providecommand \bibitemStop [0]{}%
\providecommand \bibitemNoStop [0]{.\EOS\space}%
\providecommand \EOS [0]{\spacefactor3000\relax}%
\providecommand \BibitemShut  [1]{\csname bibitem#1\endcsname}%
\let\auto@bib@innerbib\@empty
\bibitem [{\citenamefont {Bell}(1964)}]{bell1964einstein}%
  \BibitemOpen
  \bibfield  {author} {\bibinfo {author} {\bibfnamefont {J.~S.}\ \bibnamefont
  {Bell}},\ }\href {https://doi.org/10.1103/PhysicsPhysiqueFizika.1.195}
  {\bibfield  {journal} {\bibinfo  {journal} {Physics Physique Fizika}\
  }\textbf {\bibinfo {volume} {1}},\ \bibinfo {pages} {195} (\bibinfo {year}
  {1964})}\BibitemShut {NoStop}%
\bibitem [{\citenamefont {Brunner}\ \emph {et~al.}(2014)\citenamefont
  {Brunner}, \citenamefont {Cavalcanti}, \citenamefont {Pironio}, \citenamefont
  {Scarani},\ and\ \citenamefont {Wehner}}]{brunner2014bell}%
  \BibitemOpen
  \bibfield  {author} {\bibinfo {author} {\bibfnamefont {N.}~\bibnamefont
  {Brunner}}, \bibinfo {author} {\bibfnamefont {D.}~\bibnamefont {Cavalcanti}},
  \bibinfo {author} {\bibfnamefont {S.}~\bibnamefont {Pironio}}, \bibinfo
  {author} {\bibfnamefont {V.}~\bibnamefont {Scarani}},\ and\ \bibinfo {author}
  {\bibfnamefont {S.}~\bibnamefont {Wehner}},\ }\href
  {https://doi.org/10.1103/RevModPhys.86.419} {\bibfield  {journal} {\bibinfo
  {journal} {Rev. Mod. Phys.}\ }\textbf {\bibinfo {volume} {86}},\ \bibinfo
  {pages} {419} (\bibinfo {year} {2014})}\BibitemShut {NoStop}%
\bibitem [{\citenamefont {Shalm}\ \emph {et~al.}(2015)\citenamefont {Shalm},
  \citenamefont {Meyer-Scott}, \citenamefont {Christensen}, \citenamefont
  {Bierhorst}, \citenamefont {Wayne}, \citenamefont {Stevens}, \citenamefont
  {Gerrits}, \citenamefont {Glancy}, \citenamefont {Hamel}, \citenamefont
  {Allman}, \citenamefont {Coakley}, \citenamefont {Dyer}, \citenamefont
  {Hodge}, \citenamefont {Lita}, \citenamefont {Verma}, \citenamefont
  {Lambrocco}, \citenamefont {Tortorici}, \citenamefont {Migdall},
  \citenamefont {Zhang}, \citenamefont {Kumor}, \citenamefont {Farr},
  \citenamefont {Marsili}, \citenamefont {Shaw}, \citenamefont {Stern},
  \citenamefont {Abell\'an}, \citenamefont {Amaya}, \citenamefont {Pruneri},
  \citenamefont {Jennewein}, \citenamefont {Mitchell}, \citenamefont {Kwiat},
  \citenamefont {Bienfang}, \citenamefont {Mirin}, \citenamefont {Knill},\ and\
  \citenamefont {Nam}}]{shalm2015strong}%
  \BibitemOpen
  \bibfield  {author} {\bibinfo {author} {\bibfnamefont {L.~K.}\ \bibnamefont
  {Shalm}}, \bibinfo {author} {\bibfnamefont {E.}~\bibnamefont {Meyer-Scott}},
  \bibinfo {author} {\bibfnamefont {B.~G.}\ \bibnamefont {Christensen}},
  \bibinfo {author} {\bibfnamefont {P.}~\bibnamefont {Bierhorst}}, \bibinfo
  {author} {\bibfnamefont {M.~A.}\ \bibnamefont {Wayne}}, \bibinfo {author}
  {\bibfnamefont {M.~J.}\ \bibnamefont {Stevens}}, \bibinfo {author}
  {\bibfnamefont {T.}~\bibnamefont {Gerrits}}, \bibinfo {author} {\bibfnamefont
  {S.}~\bibnamefont {Glancy}}, \bibinfo {author} {\bibfnamefont {D.~R.}\
  \bibnamefont {Hamel}}, \bibinfo {author} {\bibfnamefont {M.~S.}\ \bibnamefont
  {Allman}}, \bibinfo {author} {\bibfnamefont {K.~J.}\ \bibnamefont {Coakley}},
  \bibinfo {author} {\bibfnamefont {S.~D.}\ \bibnamefont {Dyer}}, \bibinfo
  {author} {\bibfnamefont {C.}~\bibnamefont {Hodge}}, \bibinfo {author}
  {\bibfnamefont {A.~E.}\ \bibnamefont {Lita}}, \bibinfo {author}
  {\bibfnamefont {V.~B.}\ \bibnamefont {Verma}}, \bibinfo {author}
  {\bibfnamefont {C.}~\bibnamefont {Lambrocco}}, \bibinfo {author}
  {\bibfnamefont {E.}~\bibnamefont {Tortorici}}, \bibinfo {author}
  {\bibfnamefont {A.~L.}\ \bibnamefont {Migdall}}, \bibinfo {author}
  {\bibfnamefont {Y.}~\bibnamefont {Zhang}}, \bibinfo {author} {\bibfnamefont
  {D.~R.}\ \bibnamefont {Kumor}}, \bibinfo {author} {\bibfnamefont {W.~H.}\
  \bibnamefont {Farr}}, \bibinfo {author} {\bibfnamefont {F.}~\bibnamefont
  {Marsili}}, \bibinfo {author} {\bibfnamefont {M.~D.}\ \bibnamefont {Shaw}},
  \bibinfo {author} {\bibfnamefont {J.~A.}\ \bibnamefont {Stern}}, \bibinfo
  {author} {\bibfnamefont {C.}~\bibnamefont {Abell\'an}}, \bibinfo {author}
  {\bibfnamefont {W.}~\bibnamefont {Amaya}}, \bibinfo {author} {\bibfnamefont
  {V.}~\bibnamefont {Pruneri}}, \bibinfo {author} {\bibfnamefont
  {T.}~\bibnamefont {Jennewein}}, \bibinfo {author} {\bibfnamefont {M.~W.}\
  \bibnamefont {Mitchell}}, \bibinfo {author} {\bibfnamefont {P.~G.}\
  \bibnamefont {Kwiat}}, \bibinfo {author} {\bibfnamefont {J.~C.}\ \bibnamefont
  {Bienfang}}, \bibinfo {author} {\bibfnamefont {R.~P.}\ \bibnamefont {Mirin}},
  \bibinfo {author} {\bibfnamefont {E.}~\bibnamefont {Knill}},\ and\ \bibinfo
  {author} {\bibfnamefont {S.~W.}\ \bibnamefont {Nam}},\ }\href
  {https://doi.org/10.1103/PhysRevLett.115.250402} {\bibfield  {journal}
  {\bibinfo  {journal} {Phys. Rev. Lett.}\ }\textbf {\bibinfo {volume} {115}},\
  \bibinfo {pages} {250402} (\bibinfo {year} {2015})}\BibitemShut {NoStop}%
\bibitem [{\citenamefont {Giustina}\ \emph {et~al.}(2015)\citenamefont
  {Giustina}, \citenamefont {Versteegh}, \citenamefont {Wengerowsky},
  \citenamefont {Handsteiner}, \citenamefont {Hochrainer}, \citenamefont
  {Phelan}, \citenamefont {Steinlechner}, \citenamefont {Kofler}, \citenamefont
  {Larsson}, \citenamefont {Abell\'an}, \citenamefont {Amaya}, \citenamefont
  {Pruneri}, \citenamefont {Mitchell}, \citenamefont {Beyer}, \citenamefont
  {Gerrits}, \citenamefont {Lita}, \citenamefont {Shalm}, \citenamefont {Nam},
  \citenamefont {Scheidl}, \citenamefont {Ursin}, \citenamefont {Wittmann},\
  and\ \citenamefont {Zeilinger}}]{giustina2015significant}%
  \BibitemOpen
  \bibfield  {author} {\bibinfo {author} {\bibfnamefont {M.}~\bibnamefont
  {Giustina}}, \bibinfo {author} {\bibfnamefont {M.~A.~M.}\ \bibnamefont
  {Versteegh}}, \bibinfo {author} {\bibfnamefont {S.}~\bibnamefont
  {Wengerowsky}}, \bibinfo {author} {\bibfnamefont {J.}~\bibnamefont
  {Handsteiner}}, \bibinfo {author} {\bibfnamefont {A.}~\bibnamefont
  {Hochrainer}}, \bibinfo {author} {\bibfnamefont {K.}~\bibnamefont {Phelan}},
  \bibinfo {author} {\bibfnamefont {F.}~\bibnamefont {Steinlechner}}, \bibinfo
  {author} {\bibfnamefont {J.}~\bibnamefont {Kofler}}, \bibinfo {author}
  {\bibfnamefont {J.-A.}\ \bibnamefont {Larsson}}, \bibinfo {author}
  {\bibfnamefont {C.}~\bibnamefont {Abell\'an}}, \bibinfo {author}
  {\bibfnamefont {W.}~\bibnamefont {Amaya}}, \bibinfo {author} {\bibfnamefont
  {V.}~\bibnamefont {Pruneri}}, \bibinfo {author} {\bibfnamefont {M.~W.}\
  \bibnamefont {Mitchell}}, \bibinfo {author} {\bibfnamefont {J.}~\bibnamefont
  {Beyer}}, \bibinfo {author} {\bibfnamefont {T.}~\bibnamefont {Gerrits}},
  \bibinfo {author} {\bibfnamefont {A.~E.}\ \bibnamefont {Lita}}, \bibinfo
  {author} {\bibfnamefont {L.~K.}\ \bibnamefont {Shalm}}, \bibinfo {author}
  {\bibfnamefont {S.~W.}\ \bibnamefont {Nam}}, \bibinfo {author} {\bibfnamefont
  {T.}~\bibnamefont {Scheidl}}, \bibinfo {author} {\bibfnamefont
  {R.}~\bibnamefont {Ursin}}, \bibinfo {author} {\bibfnamefont
  {B.}~\bibnamefont {Wittmann}},\ and\ \bibinfo {author} {\bibfnamefont
  {A.}~\bibnamefont {Zeilinger}},\ }\href
  {https://doi.org/10.1103/PhysRevLett.115.250401} {\bibfield  {journal}
  {\bibinfo  {journal} {Phys. Rev. Lett.}\ }\textbf {\bibinfo {volume} {115}},\
  \bibinfo {pages} {250401} (\bibinfo {year} {2015})}\BibitemShut {NoStop}%
\bibitem [{\citenamefont {Hensen}\ \emph {et~al.}(2015)\citenamefont {Hensen},
  \citenamefont {Bernien}, \citenamefont {Dr{\'e}au}, \citenamefont {Reiserer},
  \citenamefont {Kalb}, \citenamefont {Blok}, \citenamefont {Ruitenberg},
  \citenamefont {Vermeulen}, \citenamefont {Schouten}, \citenamefont
  {Abell{\'a}n}, \citenamefont {Amaya}, \citenamefont {Pruneri}, \citenamefont
  {Mitchell}, \citenamefont {Markham}, \citenamefont {Twitchen}, \citenamefont
  {Elkouss}, \citenamefont {Wehner}, \citenamefont {Taminiau},\ and\
  \citenamefont {Hanson}}]{hensen2015loophole}%
  \BibitemOpen
  \bibfield  {author} {\bibinfo {author} {\bibfnamefont {B.}~\bibnamefont
  {Hensen}}, \bibinfo {author} {\bibfnamefont {H.}~\bibnamefont {Bernien}},
  \bibinfo {author} {\bibfnamefont {A.~E.}\ \bibnamefont {Dr{\'e}au}}, \bibinfo
  {author} {\bibfnamefont {A.}~\bibnamefont {Reiserer}}, \bibinfo {author}
  {\bibfnamefont {N.}~\bibnamefont {Kalb}}, \bibinfo {author} {\bibfnamefont
  {M.~S.}\ \bibnamefont {Blok}}, \bibinfo {author} {\bibfnamefont
  {J.}~\bibnamefont {Ruitenberg}}, \bibinfo {author} {\bibfnamefont {R.~F.~L.}\
  \bibnamefont {Vermeulen}}, \bibinfo {author} {\bibfnamefont {R.~N.}\
  \bibnamefont {Schouten}}, \bibinfo {author} {\bibfnamefont {C.}~\bibnamefont
  {Abell{\'a}n}}, \bibinfo {author} {\bibfnamefont {W.}~\bibnamefont {Amaya}},
  \bibinfo {author} {\bibfnamefont {V.}~\bibnamefont {Pruneri}}, \bibinfo
  {author} {\bibfnamefont {M.~W.}\ \bibnamefont {Mitchell}}, \bibinfo {author}
  {\bibfnamefont {M.}~\bibnamefont {Markham}}, \bibinfo {author} {\bibfnamefont
  {D.~J.}\ \bibnamefont {Twitchen}}, \bibinfo {author} {\bibfnamefont
  {D.}~\bibnamefont {Elkouss}}, \bibinfo {author} {\bibfnamefont
  {S.}~\bibnamefont {Wehner}}, \bibinfo {author} {\bibfnamefont {T.~H.}\
  \bibnamefont {Taminiau}},\ and\ \bibinfo {author} {\bibfnamefont
  {R.}~\bibnamefont {Hanson}},\ }\href {https://doi.org/10.1038/nature15759}
  {\bibfield  {journal} {\bibinfo  {journal} {Nature}\ }\textbf {\bibinfo
  {volume} {526}},\ \bibinfo {pages} {682} (\bibinfo {year}
  {2015})}\BibitemShut {NoStop}%
\bibitem [{\citenamefont {Rosenfeld}\ \emph {et~al.}(2017)\citenamefont
  {Rosenfeld}, \citenamefont {Burchardt}, \citenamefont {Garthoff},
  \citenamefont {Redeker}, \citenamefont {Ortegel}, \citenamefont {Rau},\ and\
  \citenamefont {Weinfurter}}]{rosenfeld2017event}%
  \BibitemOpen
  \bibfield  {author} {\bibinfo {author} {\bibfnamefont {W.}~\bibnamefont
  {Rosenfeld}}, \bibinfo {author} {\bibfnamefont {D.}~\bibnamefont
  {Burchardt}}, \bibinfo {author} {\bibfnamefont {R.}~\bibnamefont {Garthoff}},
  \bibinfo {author} {\bibfnamefont {K.}~\bibnamefont {Redeker}}, \bibinfo
  {author} {\bibfnamefont {N.}~\bibnamefont {Ortegel}}, \bibinfo {author}
  {\bibfnamefont {M.}~\bibnamefont {Rau}},\ and\ \bibinfo {author}
  {\bibfnamefont {H.}~\bibnamefont {Weinfurter}},\ }\href
  {https://doi.org/10.1103/PhysRevLett.119.010402} {\bibfield  {journal}
  {\bibinfo  {journal} {Phys. Rev. Lett.}\ }\textbf {\bibinfo {volume} {119}},\
  \bibinfo {pages} {010402} (\bibinfo {year} {2017})}\BibitemShut {NoStop}%
\bibitem [{\citenamefont {Li}\ \emph {et~al.}(2018)\citenamefont {Li},
  \citenamefont {Wu}, \citenamefont {Zhang}, \citenamefont {Liu}, \citenamefont
  {Bai}, \citenamefont {Liu}, \citenamefont {Zhang}, \citenamefont {Zhao},
  \citenamefont {Li}, \citenamefont {Wang}, \citenamefont {You}, \citenamefont
  {Munro}, \citenamefont {Yin}, \citenamefont {Zhang}, \citenamefont {Peng},
  \citenamefont {Ma}, \citenamefont {Zhang}, \citenamefont {Fan},\ and\
  \citenamefont {Pan}}]{li2018test}%
  \BibitemOpen
  \bibfield  {author} {\bibinfo {author} {\bibfnamefont {M.-H.}\ \bibnamefont
  {Li}}, \bibinfo {author} {\bibfnamefont {C.}~\bibnamefont {Wu}}, \bibinfo
  {author} {\bibfnamefont {Y.}~\bibnamefont {Zhang}}, \bibinfo {author}
  {\bibfnamefont {W.-Z.}\ \bibnamefont {Liu}}, \bibinfo {author} {\bibfnamefont
  {B.}~\bibnamefont {Bai}}, \bibinfo {author} {\bibfnamefont {Y.}~\bibnamefont
  {Liu}}, \bibinfo {author} {\bibfnamefont {W.}~\bibnamefont {Zhang}}, \bibinfo
  {author} {\bibfnamefont {Q.}~\bibnamefont {Zhao}}, \bibinfo {author}
  {\bibfnamefont {H.}~\bibnamefont {Li}}, \bibinfo {author} {\bibfnamefont
  {Z.}~\bibnamefont {Wang}}, \bibinfo {author} {\bibfnamefont {L.}~\bibnamefont
  {You}}, \bibinfo {author} {\bibfnamefont {W.~J.}\ \bibnamefont {Munro}},
  \bibinfo {author} {\bibfnamefont {J.}~\bibnamefont {Yin}}, \bibinfo {author}
  {\bibfnamefont {J.}~\bibnamefont {Zhang}}, \bibinfo {author} {\bibfnamefont
  {C.-Z.}\ \bibnamefont {Peng}}, \bibinfo {author} {\bibfnamefont
  {X.}~\bibnamefont {Ma}}, \bibinfo {author} {\bibfnamefont {Q.}~\bibnamefont
  {Zhang}}, \bibinfo {author} {\bibfnamefont {J.}~\bibnamefont {Fan}},\ and\
  \bibinfo {author} {\bibfnamefont {J.-W.}\ \bibnamefont {Pan}},\ }\href
  {https://doi.org/10.1103/PhysRevLett.121.080404} {\bibfield  {journal}
  {\bibinfo  {journal} {Phys. Rev. Lett.}\ }\textbf {\bibinfo {volume} {121}},\
  \bibinfo {pages} {080404} (\bibinfo {year} {2018})}\BibitemShut {NoStop}%
\bibitem [{\citenamefont {Clauser}\ \emph {et~al.}(1969)\citenamefont
  {Clauser}, \citenamefont {Horne}, \citenamefont {Shimony},\ and\
  \citenamefont {Holt}}]{clauser1969proposed}%
  \BibitemOpen
  \bibfield  {author} {\bibinfo {author} {\bibfnamefont {J.~F.}\ \bibnamefont
  {Clauser}}, \bibinfo {author} {\bibfnamefont {M.~A.}\ \bibnamefont {Horne}},
  \bibinfo {author} {\bibfnamefont {A.}~\bibnamefont {Shimony}},\ and\ \bibinfo
  {author} {\bibfnamefont {R.~A.}\ \bibnamefont {Holt}},\ }\href
  {https://doi.org/10.1103/PhysRevLett.23.880} {\bibfield  {journal} {\bibinfo
  {journal} {Phys. Rev. Lett.}\ }\textbf {\bibinfo {volume} {23}},\ \bibinfo
  {pages} {880} (\bibinfo {year} {1969})}\BibitemShut {NoStop}%
\bibitem [{\citenamefont {Ac\'{\i}n}\ \emph {et~al.}(2007)\citenamefont
  {Ac\'{\i}n}, \citenamefont {Brunner}, \citenamefont {Gisin}, \citenamefont
  {Massar}, \citenamefont {Pironio},\ and\ \citenamefont
  {Scarani}}]{acin2007device}%
  \BibitemOpen
  \bibfield  {author} {\bibinfo {author} {\bibfnamefont {A.}~\bibnamefont
  {Ac\'{\i}n}}, \bibinfo {author} {\bibfnamefont {N.}~\bibnamefont {Brunner}},
  \bibinfo {author} {\bibfnamefont {N.}~\bibnamefont {Gisin}}, \bibinfo
  {author} {\bibfnamefont {S.}~\bibnamefont {Massar}}, \bibinfo {author}
  {\bibfnamefont {S.}~\bibnamefont {Pironio}},\ and\ \bibinfo {author}
  {\bibfnamefont {V.}~\bibnamefont {Scarani}},\ }\href
  {https://doi.org/10.1103/PhysRevLett.98.230501} {\bibfield  {journal}
  {\bibinfo  {journal} {Phys. Rev. Lett.}\ }\textbf {\bibinfo {volume} {98}},\
  \bibinfo {pages} {230501} (\bibinfo {year} {2007})}\BibitemShut {NoStop}%
\bibitem [{\citenamefont {Ac{\'\i}n}\ and\ \citenamefont
  {Masanes}(2016)}]{acin2016certified}%
  \BibitemOpen
  \bibfield  {author} {\bibinfo {author} {\bibfnamefont {A.}~\bibnamefont
  {Ac{\'\i}n}}\ and\ \bibinfo {author} {\bibfnamefont {L.}~\bibnamefont
  {Masanes}},\ }\href {https://doi.org/10.1038/nature20119} {\bibfield
  {journal} {\bibinfo  {journal} {Nature}\ }\textbf {\bibinfo {volume} {540}},\
  \bibinfo {pages} {213} (\bibinfo {year} {2016})}\BibitemShut {NoStop}%
\bibitem [{\citenamefont {Pironio}\ \emph {et~al.}(2010)\citenamefont
  {Pironio}, \citenamefont {Ac{\'\i}n}, \citenamefont {Massar}, \citenamefont
  {Boyer~de La~Giroday}, \citenamefont {Matsukevich}, \citenamefont {Maunz},
  \citenamefont {Olmschenk}, \citenamefont {Hayes}, \citenamefont {Luo},
  \citenamefont {Manning},\ and\ \citenamefont {Monroe}}]{pironio2010random}%
  \BibitemOpen
  \bibfield  {author} {\bibinfo {author} {\bibfnamefont {S.}~\bibnamefont
  {Pironio}}, \bibinfo {author} {\bibfnamefont {A.}~\bibnamefont {Ac{\'\i}n}},
  \bibinfo {author} {\bibfnamefont {S.}~\bibnamefont {Massar}}, \bibinfo
  {author} {\bibfnamefont {A.}~\bibnamefont {Boyer~de La~Giroday}}, \bibinfo
  {author} {\bibfnamefont {D.~N.}\ \bibnamefont {Matsukevich}}, \bibinfo
  {author} {\bibfnamefont {P.}~\bibnamefont {Maunz}}, \bibinfo {author}
  {\bibfnamefont {S.}~\bibnamefont {Olmschenk}}, \bibinfo {author}
  {\bibfnamefont {D.}~\bibnamefont {Hayes}}, \bibinfo {author} {\bibfnamefont
  {L.}~\bibnamefont {Luo}}, \bibinfo {author} {\bibfnamefont {T.~A.}\
  \bibnamefont {Manning}},\ and\ \bibinfo {author} {\bibfnamefont
  {C.}~\bibnamefont {Monroe}},\ }\href {https://doi.org/10.1038/nature09008}
  {\bibfield  {journal} {\bibinfo  {journal} {Nature}\ }\textbf {\bibinfo
  {volume} {464}},\ \bibinfo {pages} {1021} (\bibinfo {year}
  {2010})}\BibitemShut {NoStop}%
\bibitem [{\citenamefont {Tavakoli}\ \emph {et~al.}(2022)\citenamefont
  {Tavakoli}, \citenamefont {Pozas-Kerstjens}, \citenamefont {Luo},\ and\
  \citenamefont {Renou}}]{tavakoli2021bell}%
  \BibitemOpen
  \bibfield  {author} {\bibinfo {author} {\bibfnamefont {A.}~\bibnamefont
  {Tavakoli}}, \bibinfo {author} {\bibfnamefont {A.}~\bibnamefont
  {Pozas-Kerstjens}}, \bibinfo {author} {\bibfnamefont {M.-X.}\ \bibnamefont
  {Luo}},\ and\ \bibinfo {author} {\bibfnamefont {M.-O.}\ \bibnamefont
  {Renou}},\ }\href {https://doi.org/10.1088/1361-6633/ac41bb} {\bibfield
  {journal} {\bibinfo  {journal} {Rep. Prog. Phys.}\ }\textbf {\bibinfo
  {volume} {85}},\ \bibinfo {pages} {056001} (\bibinfo {year}
  {2022})}\BibitemShut {NoStop}%
\bibitem [{\citenamefont {Pan}\ \emph {et~al.}(1998)\citenamefont {Pan},
  \citenamefont {Bouwmeester}, \citenamefont {Weinfurter},\ and\ \citenamefont
  {Zeilinger}}]{pan1998experimental}%
  \BibitemOpen
  \bibfield  {author} {\bibinfo {author} {\bibfnamefont {J.-W.}\ \bibnamefont
  {Pan}}, \bibinfo {author} {\bibfnamefont {D.}~\bibnamefont {Bouwmeester}},
  \bibinfo {author} {\bibfnamefont {H.}~\bibnamefont {Weinfurter}},\ and\
  \bibinfo {author} {\bibfnamefont {A.}~\bibnamefont {Zeilinger}},\ }\href
  {https://doi.org/10.1103/PhysRevLett.80.3891} {\bibfield  {journal} {\bibinfo
   {journal} {Phys. Rev. Lett.}\ }\textbf {\bibinfo {volume} {80}},\ \bibinfo
  {pages} {3891} (\bibinfo {year} {1998})}\BibitemShut {NoStop}%
\bibitem [{\citenamefont {Branciard}\ \emph {et~al.}(2010)\citenamefont
  {Branciard}, \citenamefont {Gisin},\ and\ \citenamefont
  {Pironio}}]{branciard2010characterizing}%
  \BibitemOpen
  \bibfield  {author} {\bibinfo {author} {\bibfnamefont {C.}~\bibnamefont
  {Branciard}}, \bibinfo {author} {\bibfnamefont {N.}~\bibnamefont {Gisin}},\
  and\ \bibinfo {author} {\bibfnamefont {S.}~\bibnamefont {Pironio}},\ }\href
  {https://doi.org/10.1103/PhysRevLett.104.170401} {\bibfield  {journal}
  {\bibinfo  {journal} {Phys. Rev. Lett.}\ }\textbf {\bibinfo {volume} {104}},\
  \bibinfo {pages} {170401} (\bibinfo {year} {2010})}\BibitemShut {NoStop}%
\bibitem [{\citenamefont {Branciard}\ \emph {et~al.}(2012)\citenamefont
  {Branciard}, \citenamefont {Rosset}, \citenamefont {Gisin},\ and\
  \citenamefont {Pironio}}]{branciard2012bilocal}%
  \BibitemOpen
  \bibfield  {author} {\bibinfo {author} {\bibfnamefont {C.}~\bibnamefont
  {Branciard}}, \bibinfo {author} {\bibfnamefont {D.}~\bibnamefont {Rosset}},
  \bibinfo {author} {\bibfnamefont {N.}~\bibnamefont {Gisin}},\ and\ \bibinfo
  {author} {\bibfnamefont {S.}~\bibnamefont {Pironio}},\ }\href
  {https://doi.org/10.1103/PhysRevA.85.032119} {\bibfield  {journal} {\bibinfo
  {journal} {Phys. Rev. A}\ }\textbf {\bibinfo {volume} {85}},\ \bibinfo
  {pages} {032119} (\bibinfo {year} {2012})}\BibitemShut {NoStop}%
\bibitem [{\citenamefont {Carvacho}\ \emph {et~al.}(2017)\citenamefont
  {Carvacho}, \citenamefont {Andreoli}, \citenamefont {Santodonato},
  \citenamefont {Bentivegna}, \citenamefont {Chaves},\ and\ \citenamefont
  {Sciarrino}}]{carvacho2017experimental}%
  \BibitemOpen
  \bibfield  {author} {\bibinfo {author} {\bibfnamefont {G.}~\bibnamefont
  {Carvacho}}, \bibinfo {author} {\bibfnamefont {F.}~\bibnamefont {Andreoli}},
  \bibinfo {author} {\bibfnamefont {L.}~\bibnamefont {Santodonato}}, \bibinfo
  {author} {\bibfnamefont {M.}~\bibnamefont {Bentivegna}}, \bibinfo {author}
  {\bibfnamefont {R.}~\bibnamefont {Chaves}},\ and\ \bibinfo {author}
  {\bibfnamefont {F.}~\bibnamefont {Sciarrino}},\ }\href
  {https://doi.org/10.1038/ncomms14775} {\bibfield  {journal} {\bibinfo
  {journal} {Nat. Commun.}\ }\textbf {\bibinfo {volume} {8}},\ \bibinfo {pages}
  {1} (\bibinfo {year} {2017})}\BibitemShut {NoStop}%
\bibitem [{\citenamefont {Saunders}\ \emph {et~al.}(2017)\citenamefont
  {Saunders}, \citenamefont {Bennet}, \citenamefont {Branciard},\ and\
  \citenamefont {Pryde}}]{saunders2017experimental}%
  \BibitemOpen
  \bibfield  {author} {\bibinfo {author} {\bibfnamefont {D.~J.}\ \bibnamefont
  {Saunders}}, \bibinfo {author} {\bibfnamefont {A.~J.}\ \bibnamefont
  {Bennet}}, \bibinfo {author} {\bibfnamefont {C.}~\bibnamefont {Branciard}},\
  and\ \bibinfo {author} {\bibfnamefont {G.~J.}\ \bibnamefont {Pryde}},\ }\href
  {https://doi.org/10.1126/sciadv.1602743} {\bibfield  {journal} {\bibinfo
  {journal} {Sci. Adv.}\ }\textbf {\bibinfo {volume} {3}},\ \bibinfo {pages}
  {e1602743} (\bibinfo {year} {2017})}\BibitemShut {NoStop}%
\bibitem [{\citenamefont {Andreoli}\ \emph
  {et~al.}(2017{\natexlab{a}})\citenamefont {Andreoli}, \citenamefont
  {Carvacho}, \citenamefont {Santodonato}, \citenamefont {Bentivegna},
  \citenamefont {Chaves},\ and\ \citenamefont
  {Sciarrino}}]{andreoli2017experimental}%
  \BibitemOpen
  \bibfield  {author} {\bibinfo {author} {\bibfnamefont {F.}~\bibnamefont
  {Andreoli}}, \bibinfo {author} {\bibfnamefont {G.}~\bibnamefont {Carvacho}},
  \bibinfo {author} {\bibfnamefont {L.}~\bibnamefont {Santodonato}}, \bibinfo
  {author} {\bibfnamefont {M.}~\bibnamefont {Bentivegna}}, \bibinfo {author}
  {\bibfnamefont {R.}~\bibnamefont {Chaves}},\ and\ \bibinfo {author}
  {\bibfnamefont {F.}~\bibnamefont {Sciarrino}},\ }\href
  {https://doi.org/10.1103/PhysRevA.95.062315} {\bibfield  {journal} {\bibinfo
  {journal} {Phys. Rev. A}\ }\textbf {\bibinfo {volume} {95}},\ \bibinfo
  {pages} {062315} (\bibinfo {year} {2017}{\natexlab{a}})}\BibitemShut
  {NoStop}%
\bibitem [{\citenamefont {Sun}\ \emph {et~al.}(2019)\citenamefont {Sun},
  \citenamefont {Jiang}, \citenamefont {Bai}, \citenamefont {Zhang},
  \citenamefont {Li}, \citenamefont {Jiang}, \citenamefont {Zhang},
  \citenamefont {You}, \citenamefont {Chen}, \citenamefont {Wang},
  \citenamefont {Zhang}, \citenamefont {Fan},\ and\ \citenamefont
  {Pan}}]{sun2019experimental}%
  \BibitemOpen
  \bibfield  {author} {\bibinfo {author} {\bibfnamefont {Q.-C.}\ \bibnamefont
  {Sun}}, \bibinfo {author} {\bibfnamefont {Y.-F.}\ \bibnamefont {Jiang}},
  \bibinfo {author} {\bibfnamefont {B.}~\bibnamefont {Bai}}, \bibinfo {author}
  {\bibfnamefont {W.}~\bibnamefont {Zhang}}, \bibinfo {author} {\bibfnamefont
  {H.}~\bibnamefont {Li}}, \bibinfo {author} {\bibfnamefont {X.}~\bibnamefont
  {Jiang}}, \bibinfo {author} {\bibfnamefont {J.}~\bibnamefont {Zhang}},
  \bibinfo {author} {\bibfnamefont {L.}~\bibnamefont {You}}, \bibinfo {author}
  {\bibfnamefont {X.}~\bibnamefont {Chen}}, \bibinfo {author} {\bibfnamefont
  {Z.}~\bibnamefont {Wang}}, \bibinfo {author} {\bibfnamefont {Q.}~\bibnamefont
  {Zhang}}, \bibinfo {author} {\bibfnamefont {J.}~\bibnamefont {Fan}},\ and\
  \bibinfo {author} {\bibfnamefont {J.-W.}\ \bibnamefont {Pan}},\ }\href
  {https://doi.org/10.1038/s41566-019-0502-7} {\bibfield  {journal} {\bibinfo
  {journal} {Nat. Photon.}\ }\textbf {\bibinfo {volume} {13}},\ \bibinfo
  {pages} {687} (\bibinfo {year} {2019})}\BibitemShut {NoStop}%
\bibitem [{\citenamefont {Poderini}\ \emph {et~al.}(2020)\citenamefont
  {Poderini}, \citenamefont {Agresti}, \citenamefont {Marchese}, \citenamefont
  {Polino}, \citenamefont {Giordani}, \citenamefont {Suprano}, \citenamefont
  {Valeri}, \citenamefont {Milani}, \citenamefont {Spagnolo}, \citenamefont
  {Carvacho}, \citenamefont {Chaves},\ and\ \citenamefont
  {Sciarrino}}]{poderini2020experimental}%
  \BibitemOpen
  \bibfield  {author} {\bibinfo {author} {\bibfnamefont {D.}~\bibnamefont
  {Poderini}}, \bibinfo {author} {\bibfnamefont {I.}~\bibnamefont {Agresti}},
  \bibinfo {author} {\bibfnamefont {G.}~\bibnamefont {Marchese}}, \bibinfo
  {author} {\bibfnamefont {E.}~\bibnamefont {Polino}}, \bibinfo {author}
  {\bibfnamefont {T.}~\bibnamefont {Giordani}}, \bibinfo {author}
  {\bibfnamefont {A.}~\bibnamefont {Suprano}}, \bibinfo {author} {\bibfnamefont
  {M.}~\bibnamefont {Valeri}}, \bibinfo {author} {\bibfnamefont
  {G.}~\bibnamefont {Milani}}, \bibinfo {author} {\bibfnamefont
  {N.}~\bibnamefont {Spagnolo}}, \bibinfo {author} {\bibfnamefont
  {G.}~\bibnamefont {Carvacho}}, \bibinfo {author} {\bibfnamefont
  {R.}~\bibnamefont {Chaves}},\ and\ \bibinfo {author} {\bibfnamefont
  {F.}~\bibnamefont {Sciarrino}},\ }\href
  {https://doi.org/10.1038/s41467-020-16189-6} {\bibfield  {journal} {\bibinfo
  {journal} {Nat. Commun.}\ }\textbf {\bibinfo {volume} {11}},\ \bibinfo
  {pages} {1} (\bibinfo {year} {2020})}\BibitemShut {NoStop}%
\bibitem [{\citenamefont {Pozas-Kerstjens}\ \emph {et~al.}(2022)\citenamefont
  {Pozas-Kerstjens}, \citenamefont {Gisin},\ and\ \citenamefont
  {Tavakoli}}]{pozas2022full}%
  \BibitemOpen
  \bibfield  {author} {\bibinfo {author} {\bibfnamefont {A.}~\bibnamefont
  {Pozas-Kerstjens}}, \bibinfo {author} {\bibfnamefont {N.}~\bibnamefont
  {Gisin}},\ and\ \bibinfo {author} {\bibfnamefont {A.}~\bibnamefont
  {Tavakoli}},\ }\href {https://doi.org/10.1103/PhysRevLett.128.010403}
  {\bibfield  {journal} {\bibinfo  {journal} {Phys. Rev. Lett.}\ }\textbf
  {\bibinfo {volume} {128}},\ \bibinfo {pages} {010403} (\bibinfo {year}
  {2022})}\BibitemShut {NoStop}%
\bibitem [{\citenamefont {Popescu}\ and\ \citenamefont
  {Rohrlich}(1994)}]{popescu1994quantum}%
  \BibitemOpen
  \bibfield  {author} {\bibinfo {author} {\bibfnamefont {S.}~\bibnamefont
  {Popescu}}\ and\ \bibinfo {author} {\bibfnamefont {D.}~\bibnamefont
  {Rohrlich}},\ }\href {https://doi.org/10.1007/BF02058098} {\bibfield
  {journal} {\bibinfo  {journal} {Found. Phys.}\ }\textbf {\bibinfo {volume}
  {24}},\ \bibinfo {pages} {379} (\bibinfo {year} {1994})}\BibitemShut
  {NoStop}%
\bibitem [{\citenamefont {Gisin}\ \emph {et~al.}(2020)\citenamefont {Gisin},
  \citenamefont {Bancal}, \citenamefont {Cai}, \citenamefont {Remy},
  \citenamefont {Tavakoli}, \citenamefont {Zambrini~Cruzeiro}, \citenamefont
  {Popescu},\ and\ \citenamefont {Brunner}}]{gisin2020constraints}%
  \BibitemOpen
  \bibfield  {author} {\bibinfo {author} {\bibfnamefont {N.}~\bibnamefont
  {Gisin}}, \bibinfo {author} {\bibfnamefont {J.-D.}\ \bibnamefont {Bancal}},
  \bibinfo {author} {\bibfnamefont {Y.}~\bibnamefont {Cai}}, \bibinfo {author}
  {\bibfnamefont {P.}~\bibnamefont {Remy}}, \bibinfo {author} {\bibfnamefont
  {A.}~\bibnamefont {Tavakoli}}, \bibinfo {author} {\bibfnamefont
  {E.}~\bibnamefont {Zambrini~Cruzeiro}}, \bibinfo {author} {\bibfnamefont
  {S.}~\bibnamefont {Popescu}},\ and\ \bibinfo {author} {\bibfnamefont
  {N.}~\bibnamefont {Brunner}},\ }\href
  {https://doi.org/10.1038/s41467-020-16137-4} {\bibfield  {journal} {\bibinfo
  {journal} {Nat. Commun.}\ }\textbf {\bibinfo {volume} {11}},\ \bibinfo
  {pages} {1} (\bibinfo {year} {2020})}\BibitemShut {NoStop}%
\bibitem [{\citenamefont {Coiteux-Roy}\ \emph {et~al.}(2021)\citenamefont
  {Coiteux-Roy}, \citenamefont {Wolfe},\ and\ \citenamefont
  {Renou}}]{coiteuxroy2021}%
  \BibitemOpen
  \bibfield  {author} {\bibinfo {author} {\bibfnamefont {X.}~\bibnamefont
  {Coiteux-Roy}}, \bibinfo {author} {\bibfnamefont {E.}~\bibnamefont {Wolfe}},\
  and\ \bibinfo {author} {\bibfnamefont {M.-O.}\ \bibnamefont {Renou}},\ }\href
  {https://doi.org/10.1103/PhysRevLett.127.200401} {\bibfield  {journal}
  {\bibinfo  {journal} {Phys. Rev. Lett.}\ }\textbf {\bibinfo {volume} {127}},\
  \bibinfo {pages} {200401} (\bibinfo {year} {2021})}\BibitemShut {NoStop}%
\bibitem [{\citenamefont {H{\aa}kansson}\ \emph {et~al.}(2022)\citenamefont
  {H{\aa}kansson}, \citenamefont {Piveteau}, \citenamefont {Muhammad},\ and\
  \citenamefont {Bourennane}}]{haakansson2022experimental}%
  \BibitemOpen
  \bibfield  {author} {\bibinfo {author} {\bibfnamefont {E.}~\bibnamefont
  {H{\aa}kansson}}, \bibinfo {author} {\bibfnamefont {A.}~\bibnamefont
  {Piveteau}}, \bibinfo {author} {\bibfnamefont {S.}~\bibnamefont {Muhammad}},\
  and\ \bibinfo {author} {\bibfnamefont {M.}~\bibnamefont {Bourennane}},\
  }\href {https://www.arxiv.org/abs/2201.06361} {\bibfield  {journal} {\bibinfo
   {journal} {arXiv preprint arXiv:2201.06361}\ } (\bibinfo {year}
  {2022})}\BibitemShut {NoStop}%
\bibitem [{\citenamefont {Huang}\ \emph {et~al.}(2022)\citenamefont {Huang},
  \citenamefont {Hu}, \citenamefont {Guo}, \citenamefont {Zhang}, \citenamefont
  {Liu}, \citenamefont {Huang}, \citenamefont {Li}, \citenamefont {Guo},
  \citenamefont {Gisin}, \citenamefont {Branciard},\ and\ \citenamefont
  {Tavakoli}}]{huang2022experimental}%
  \BibitemOpen
  \bibfield  {author} {\bibinfo {author} {\bibfnamefont {C.-X.}\ \bibnamefont
  {Huang}}, \bibinfo {author} {\bibfnamefont {X.-M.}\ \bibnamefont {Hu}},
  \bibinfo {author} {\bibfnamefont {Y.}~\bibnamefont {Guo}}, \bibinfo {author}
  {\bibfnamefont {C.}~\bibnamefont {Zhang}}, \bibinfo {author} {\bibfnamefont
  {B.-H.}\ \bibnamefont {Liu}}, \bibinfo {author} {\bibfnamefont {Y.-F.}\
  \bibnamefont {Huang}}, \bibinfo {author} {\bibfnamefont {C.-F.}\ \bibnamefont
  {Li}}, \bibinfo {author} {\bibfnamefont {G.-C.}\ \bibnamefont {Guo}},
  \bibinfo {author} {\bibfnamefont {N.}~\bibnamefont {Gisin}}, \bibinfo
  {author} {\bibfnamefont {C.}~\bibnamefont {Branciard}},\ and\ \bibinfo
  {author} {\bibfnamefont {A.}~\bibnamefont {Tavakoli}},\ }\href
  {https://doi.org/10.1103/PhysRevLett.129.030502} {\bibfield  {journal}
  {\bibinfo  {journal} {Phys. Rev. Lett.}\ }\textbf {\bibinfo {volume} {129}},\
  \bibinfo {pages} {030502} (\bibinfo {year} {2022})}\BibitemShut {NoStop}%
\bibitem [{\citenamefont {Wang}\ \emph {et~al.}(2023)\citenamefont {Wang},
  \citenamefont {Pozas-Kerstjens}, \citenamefont {Zhang}, \citenamefont {Liu},
  \citenamefont {Huang}, \citenamefont {Li}, \citenamefont {Guo}, \citenamefont
  {Gisin},\ and\ \citenamefont {Tavakoli}}]{wang2022experimental}%
  \BibitemOpen
  \bibfield  {author} {\bibinfo {author} {\bibfnamefont {N.-N.}\ \bibnamefont
  {Wang}}, \bibinfo {author} {\bibfnamefont {A.}~\bibnamefont
  {Pozas-Kerstjens}}, \bibinfo {author} {\bibfnamefont {C.}~\bibnamefont
  {Zhang}}, \bibinfo {author} {\bibfnamefont {B.-H.}\ \bibnamefont {Liu}},
  \bibinfo {author} {\bibfnamefont {Y.-F.}\ \bibnamefont {Huang}}, \bibinfo
  {author} {\bibfnamefont {C.-F.}\ \bibnamefont {Li}}, \bibinfo {author}
  {\bibfnamefont {G.-C.}\ \bibnamefont {Guo}}, \bibinfo {author} {\bibfnamefont
  {N.}~\bibnamefont {Gisin}},\ and\ \bibinfo {author} {\bibfnamefont
  {A.}~\bibnamefont {Tavakoli}},\ }\href
  {https://doi.org/10.1038/s41467-023-37842-w} {\bibfield  {journal} {\bibinfo
  {journal} {Nat. Commun.}\ }\textbf {\bibinfo {volume} {14}},\ \bibinfo
  {pages} {2153} (\bibinfo {year} {2023})}\BibitemShut {NoStop}%
\bibitem [{\citenamefont {Lee}\ and\ \citenamefont
  {Hoban}(2018)}]{lee2018towards}%
  \BibitemOpen
  \bibfield  {author} {\bibinfo {author} {\bibfnamefont {C.~M.}\ \bibnamefont
  {Lee}}\ and\ \bibinfo {author} {\bibfnamefont {M.~J.}\ \bibnamefont
  {Hoban}},\ }\href {https://doi.org/10.1103/PhysRevLett.120.020504} {\bibfield
   {journal} {\bibinfo  {journal} {Phys. Rev. Lett.}\ }\textbf {\bibinfo
  {volume} {120}},\ \bibinfo {pages} {020504} (\bibinfo {year}
  {2018})}\BibitemShut {NoStop}%
\bibitem [{\citenamefont {Wu}\ \emph {et~al.}(2022)\citenamefont {Wu},
  \citenamefont {Jiang}, \citenamefont {Gu}, \citenamefont {Huang},
  \citenamefont {Bai}, \citenamefont {Sun}, \citenamefont {Zhang},
  \citenamefont {Gong}, \citenamefont {Mao}, \citenamefont {Zhong},
  \citenamefont {Chen}, \citenamefont {Zhang}, \citenamefont {Zhang},
  \citenamefont {Lu},\ and\ \citenamefont {Pan}}]{wu2022experimental}%
  \BibitemOpen
  \bibfield  {author} {\bibinfo {author} {\bibfnamefont {D.}~\bibnamefont
  {Wu}}, \bibinfo {author} {\bibfnamefont {Y.-F.}\ \bibnamefont {Jiang}},
  \bibinfo {author} {\bibfnamefont {X.-M.}\ \bibnamefont {Gu}}, \bibinfo
  {author} {\bibfnamefont {L.}~\bibnamefont {Huang}}, \bibinfo {author}
  {\bibfnamefont {B.}~\bibnamefont {Bai}}, \bibinfo {author} {\bibfnamefont
  {Q.-C.}\ \bibnamefont {Sun}}, \bibinfo {author} {\bibfnamefont
  {X.}~\bibnamefont {Zhang}}, \bibinfo {author} {\bibfnamefont {S.-Q.}\
  \bibnamefont {Gong}}, \bibinfo {author} {\bibfnamefont {Y.}~\bibnamefont
  {Mao}}, \bibinfo {author} {\bibfnamefont {H.-S.}\ \bibnamefont {Zhong}},
  \bibinfo {author} {\bibfnamefont {M.-C.}\ \bibnamefont {Chen}}, \bibinfo
  {author} {\bibfnamefont {J.}~\bibnamefont {Zhang}}, \bibinfo {author}
  {\bibfnamefont {Q.}~\bibnamefont {Zhang}}, \bibinfo {author} {\bibfnamefont
  {C.-Y.}\ \bibnamefont {Lu}},\ and\ \bibinfo {author} {\bibfnamefont {J.-W.}\
  \bibnamefont {Pan}},\ }\href {https://doi.org/10.1103/PhysRevLett.129.140401}
  {\bibfield  {journal} {\bibinfo  {journal} {Phys. Rev. Lett.}\ }\textbf
  {\bibinfo {volume} {129}},\ \bibinfo {pages} {140401} (\bibinfo {year}
  {2022})}\BibitemShut {NoStop}%
\bibitem [{\citenamefont {Pan}\ and\ \citenamefont
  {Zeilinger}(1998)}]{pan1998greenberger}%
  \BibitemOpen
  \bibfield  {author} {\bibinfo {author} {\bibfnamefont {J.-w.}\ \bibnamefont
  {Pan}}\ and\ \bibinfo {author} {\bibfnamefont {A.}~\bibnamefont
  {Zeilinger}},\ }\href {https://doi.org/10.1103/PhysRevA.57.2208} {\bibfield
  {journal} {\bibinfo  {journal} {Phys. Rev. A}\ }\textbf {\bibinfo {volume}
  {57}},\ \bibinfo {pages} {2208} (\bibinfo {year} {1998})}\BibitemShut
  {NoStop}%
\bibitem [{\citenamefont {Abell\'an}\ \emph {et~al.}(2015)\citenamefont
  {Abell\'an}, \citenamefont {Amaya}, \citenamefont {Mitrani}, \citenamefont
  {Pruneri},\ and\ \citenamefont {Mitchell}}]{abellan2015generation}%
  \BibitemOpen
  \bibfield  {author} {\bibinfo {author} {\bibfnamefont {C.}~\bibnamefont
  {Abell\'an}}, \bibinfo {author} {\bibfnamefont {W.}~\bibnamefont {Amaya}},
  \bibinfo {author} {\bibfnamefont {D.}~\bibnamefont {Mitrani}}, \bibinfo
  {author} {\bibfnamefont {V.}~\bibnamefont {Pruneri}},\ and\ \bibinfo {author}
  {\bibfnamefont {M.~W.}\ \bibnamefont {Mitchell}},\ }\href
  {https://doi.org/10.1103/PhysRevLett.115.250403} {\bibfield  {journal}
  {\bibinfo  {journal} {Phys. Rev. Lett.}\ }\textbf {\bibinfo {volume} {115}},\
  \bibinfo {pages} {250403} (\bibinfo {year} {2015})}\BibitemShut {NoStop}%
\bibitem [{\citenamefont {Bassham}\ \emph {et~al.}(2010)\citenamefont
  {Bassham}, \citenamefont {Rukhin}, \citenamefont {Soto}, \citenamefont
  {Nechvatal}, \citenamefont {Smid}, \citenamefont {Barker}, \citenamefont
  {Leigh}, \citenamefont {Levenson}, \citenamefont {Vangel}, \citenamefont
  {Banks}, \citenamefont {Heckert}, \citenamefont {Dray},\ and\ \citenamefont
  {Vo}}]{rukhin2010statistical}%
  \BibitemOpen
  \bibfield  {author} {\bibinfo {author} {\bibfnamefont {L.~E.}\ \bibnamefont
  {Bassham}}, \bibinfo {author} {\bibfnamefont {A.}~\bibnamefont {Rukhin}},
  \bibinfo {author} {\bibfnamefont {J.}~\bibnamefont {Soto}}, \bibinfo {author}
  {\bibfnamefont {J.}~\bibnamefont {Nechvatal}}, \bibinfo {author}
  {\bibfnamefont {M.}~\bibnamefont {Smid}}, \bibinfo {author} {\bibfnamefont
  {E.}~\bibnamefont {Barker}}, \bibinfo {author} {\bibfnamefont
  {S.}~\bibnamefont {Leigh}}, \bibinfo {author} {\bibfnamefont
  {M.}~\bibnamefont {Levenson}}, \bibinfo {author} {\bibfnamefont
  {M.}~\bibnamefont {Vangel}}, \bibinfo {author} {\bibfnamefont
  {D.}~\bibnamefont {Banks}}, \bibinfo {author} {\bibfnamefont
  {A.}~\bibnamefont {Heckert}}, \bibinfo {author} {\bibfnamefont
  {J.}~\bibnamefont {Dray}},\ and\ \bibinfo {author} {\bibfnamefont
  {S.}~\bibnamefont {Vo}},\ }\href
  {https://tsapps.nist.gov/publication/get_pdf.cfm?pub_id=906762} {\bibinfo
  {title} {A statistical test suite for random and pseudorandom number
  generators for cryptographic applications}} (\bibinfo {year} {2010}),\
  \bibinfo {note} {{NIST} Special Publication 800-22 rev. 1a}\BibitemShut
  {NoStop}%
\bibitem [{\citenamefont {Liu}\ \emph {et~al.}(2018)\citenamefont {Liu},
  \citenamefont {Zhao}, \citenamefont {Li}, \citenamefont {Guan}, \citenamefont
  {Zhang}, \citenamefont {Bai}, \citenamefont {Zhang}, \citenamefont {Liu},
  \citenamefont {Wu}, \citenamefont {Yuan}, \citenamefont {Li}, \citenamefont
  {Munro}, \citenamefont {Wang}, \citenamefont {You}, \citenamefont {Zhang},
  \citenamefont {Ma}, \citenamefont {Fan}, \citenamefont {Zhang},\ and\
  \citenamefont {Pan}}]{liu2018device}%
  \BibitemOpen
  \bibfield  {author} {\bibinfo {author} {\bibfnamefont {Y.}~\bibnamefont
  {Liu}}, \bibinfo {author} {\bibfnamefont {Q.}~\bibnamefont {Zhao}}, \bibinfo
  {author} {\bibfnamefont {M.-H.}\ \bibnamefont {Li}}, \bibinfo {author}
  {\bibfnamefont {J.-Y.}\ \bibnamefont {Guan}}, \bibinfo {author}
  {\bibfnamefont {Y.}~\bibnamefont {Zhang}}, \bibinfo {author} {\bibfnamefont
  {B.}~\bibnamefont {Bai}}, \bibinfo {author} {\bibfnamefont {W.}~\bibnamefont
  {Zhang}}, \bibinfo {author} {\bibfnamefont {W.-Z.}\ \bibnamefont {Liu}},
  \bibinfo {author} {\bibfnamefont {C.}~\bibnamefont {Wu}}, \bibinfo {author}
  {\bibfnamefont {X.}~\bibnamefont {Yuan}}, \bibinfo {author} {\bibfnamefont
  {H.}~\bibnamefont {Li}}, \bibinfo {author} {\bibfnamefont {W.~J.}\
  \bibnamefont {Munro}}, \bibinfo {author} {\bibfnamefont {Z.}~\bibnamefont
  {Wang}}, \bibinfo {author} {\bibfnamefont {L.}~\bibnamefont {You}}, \bibinfo
  {author} {\bibfnamefont {J.}~\bibnamefont {Zhang}}, \bibinfo {author}
  {\bibfnamefont {X.}~\bibnamefont {Ma}}, \bibinfo {author} {\bibfnamefont
  {J.}~\bibnamefont {Fan}}, \bibinfo {author} {\bibfnamefont {Q.}~\bibnamefont
  {Zhang}},\ and\ \bibinfo {author} {\bibfnamefont {J.-W.}\ \bibnamefont
  {Pan}},\ }\href {https://doi.org/0.1038/s41586-018-0559-3} {\bibfield
  {journal} {\bibinfo  {journal} {Nature}\ }\textbf {\bibinfo {volume} {562}},\
  \bibinfo {pages} {548} (\bibinfo {year} {2018})}\BibitemShut {NoStop}%
\bibitem [{\citenamefont {Pearle}(1970)}]{pearle1970}%
  \BibitemOpen
  \bibfield  {author} {\bibinfo {author} {\bibfnamefont {P.~M.}\ \bibnamefont
  {Pearle}},\ }\href {https://doi.org/10.1103/PhysRevD.2.1418} {\bibfield
  {journal} {\bibinfo  {journal} {Phys. Rev. D}\ }\textbf {\bibinfo {volume}
  {2}},\ \bibinfo {pages} {1418} (\bibinfo {year} {1970})}\BibitemShut
  {NoStop}%
\bibitem [{\citenamefont {Accardi}\ and\ \citenamefont
  {Regoli}(2001)}]{accardi}%
  \BibitemOpen
  \bibfield  {author} {\bibinfo {author} {\bibfnamefont {L.}~\bibnamefont
  {Accardi}}\ and\ \bibinfo {author} {\bibfnamefont {M.}~\bibnamefont
  {Regoli}},\ }\bibinfo {title} {Locality and {B}ell's inequality},\ in\ \href
  {https://doi.org/10.1142/9789812810809_0001} {\emph {\bibinfo {booktitle}
  {Foundations of Probability and Physics}}}\ (\bibinfo  {publisher} {World
  Scientific},\ \bibinfo {year} {2001})\ pp.\ \bibinfo {pages}
  {1--28}\BibitemShut {NoStop}%
\bibitem [{\citenamefont {Barrett}\ \emph {et~al.}(2002)\citenamefont
  {Barrett}, \citenamefont {Collins}, \citenamefont {Hardy}, \citenamefont
  {Kent},\ and\ \citenamefont {Popescu}}]{barrett2002}%
  \BibitemOpen
  \bibfield  {author} {\bibinfo {author} {\bibfnamefont {J.}~\bibnamefont
  {Barrett}}, \bibinfo {author} {\bibfnamefont {D.}~\bibnamefont {Collins}},
  \bibinfo {author} {\bibfnamefont {L.}~\bibnamefont {Hardy}}, \bibinfo
  {author} {\bibfnamefont {A.}~\bibnamefont {Kent}},\ and\ \bibinfo {author}
  {\bibfnamefont {S.}~\bibnamefont {Popescu}},\ }\href
  {https://doi.org/10.1103/PhysRevA.66.042111} {\bibfield  {journal} {\bibinfo
  {journal} {Phys. Rev. A}\ }\textbf {\bibinfo {volume} {66}},\ \bibinfo
  {pages} {042111} (\bibinfo {year} {2002})}\BibitemShut {NoStop}%
\bibitem [{\citenamefont {Wang}\ \emph {et~al.}(2019)\citenamefont {Wang},
  \citenamefont {He}, \citenamefont {Chung}, \citenamefont {Hu}, \citenamefont
  {Yu}, \citenamefont {Chen}, \citenamefont {Ding}, \citenamefont {Chen},
  \citenamefont {Qin}, \citenamefont {Yang}, \citenamefont {Liu}, \citenamefont
  {Duan}, \citenamefont {Li}, \citenamefont {Gerhardt}, \citenamefont
  {Winkler}, \citenamefont {Jurkat}, \citenamefont {Wang}, \citenamefont
  {Gregersen}, \citenamefont {Huo}, \citenamefont {Dai}, \citenamefont {Yu},
  \citenamefont {H\"ofling}, \citenamefont {Lu},\ and\ \citenamefont
  {Pan}}]{wang2019towards}%
  \BibitemOpen
  \bibfield  {author} {\bibinfo {author} {\bibfnamefont {H.}~\bibnamefont
  {Wang}}, \bibinfo {author} {\bibfnamefont {Y.-M.}\ \bibnamefont {He}},
  \bibinfo {author} {\bibfnamefont {T.-H.}\ \bibnamefont {Chung}}, \bibinfo
  {author} {\bibfnamefont {H.}~\bibnamefont {Hu}}, \bibinfo {author}
  {\bibfnamefont {Y.}~\bibnamefont {Yu}}, \bibinfo {author} {\bibfnamefont
  {S.}~\bibnamefont {Chen}}, \bibinfo {author} {\bibfnamefont {X.}~\bibnamefont
  {Ding}}, \bibinfo {author} {\bibfnamefont {M.-C.}\ \bibnamefont {Chen}},
  \bibinfo {author} {\bibfnamefont {J.}~\bibnamefont {Qin}}, \bibinfo {author}
  {\bibfnamefont {X.}~\bibnamefont {Yang}}, \bibinfo {author} {\bibfnamefont
  {R.-Z.}\ \bibnamefont {Liu}}, \bibinfo {author} {\bibfnamefont {Z.-C.}\
  \bibnamefont {Duan}}, \bibinfo {author} {\bibfnamefont {J.-P.}\ \bibnamefont
  {Li}}, \bibinfo {author} {\bibfnamefont {S.}~\bibnamefont {Gerhardt}},
  \bibinfo {author} {\bibfnamefont {K.}~\bibnamefont {Winkler}}, \bibinfo
  {author} {\bibfnamefont {J.}~\bibnamefont {Jurkat}}, \bibinfo {author}
  {\bibfnamefont {L.-J.}\ \bibnamefont {Wang}}, \bibinfo {author}
  {\bibfnamefont {N.}~\bibnamefont {Gregersen}}, \bibinfo {author}
  {\bibfnamefont {Y.-H.}\ \bibnamefont {Huo}}, \bibinfo {author} {\bibfnamefont
  {Q.}~\bibnamefont {Dai}}, \bibinfo {author} {\bibfnamefont {S.}~\bibnamefont
  {Yu}}, \bibinfo {author} {\bibfnamefont {S.}~\bibnamefont {H\"ofling}},
  \bibinfo {author} {\bibfnamefont {C.-Y.}\ \bibnamefont {Lu}},\ and\ \bibinfo
  {author} {\bibfnamefont {J.-W.}\ \bibnamefont {Pan}},\ }\href
  {https://doi.org/10.1038/s41566-019-0494-3} {\bibfield  {journal} {\bibinfo
  {journal} {Nat. Photon.}\ }\textbf {\bibinfo {volume} {13}},\ \bibinfo
  {pages} {770} (\bibinfo {year} {2019})}\BibitemShut {NoStop}%
\bibitem [{\citenamefont {Gill}(2001)}]{gill2001}%
  \BibitemOpen
  \bibfield  {author} {\bibinfo {author} {\bibfnamefont {R.~D.}\ \bibnamefont
  {Gill}},\ }\href {https://arxiv.org/abs/quant-ph/0110137} {\bibfield
  {journal} {\bibinfo  {journal} {arXiv preprint arXiv:quant-ph/0110137}\ }
  (\bibinfo {year} {2001})}\BibitemShut {NoStop}%
\bibitem [{\citenamefont {Tavakoli}\ \emph {et~al.}(2014)\citenamefont
  {Tavakoli}, \citenamefont {Skrzypczyk}, \citenamefont {Cavalcanti},\ and\
  \citenamefont {Ac\'{\i}n}}]{tavakoli2014nonlocal}%
  \BibitemOpen
  \bibfield  {author} {\bibinfo {author} {\bibfnamefont {A.}~\bibnamefont
  {Tavakoli}}, \bibinfo {author} {\bibfnamefont {P.}~\bibnamefont
  {Skrzypczyk}}, \bibinfo {author} {\bibfnamefont {D.}~\bibnamefont
  {Cavalcanti}},\ and\ \bibinfo {author} {\bibfnamefont {A.}~\bibnamefont
  {Ac\'{\i}n}},\ }\href {https://doi.org/10.1103/PhysRevA.90.062109} {\bibfield
   {journal} {\bibinfo  {journal} {Phys. Rev. A}\ }\textbf {\bibinfo {volume}
  {90}},\ \bibinfo {pages} {062109} (\bibinfo {year} {2014})}\BibitemShut
  {NoStop}%
\bibitem [{\citenamefont {Andreoli}\ \emph
  {et~al.}(2017{\natexlab{b}})\citenamefont {Andreoli}, \citenamefont
  {Carvacho}, \citenamefont {Santodonato}, \citenamefont {Chaves},\ and\
  \citenamefont {Sciarrino}}]{andreoli2017maximal}%
  \BibitemOpen
  \bibfield  {author} {\bibinfo {author} {\bibfnamefont {F.}~\bibnamefont
  {Andreoli}}, \bibinfo {author} {\bibfnamefont {G.}~\bibnamefont {Carvacho}},
  \bibinfo {author} {\bibfnamefont {L.}~\bibnamefont {Santodonato}}, \bibinfo
  {author} {\bibfnamefont {R.}~\bibnamefont {Chaves}},\ and\ \bibinfo {author}
  {\bibfnamefont {F.}~\bibnamefont {Sciarrino}},\ }\href
  {https://doi.org/10.1088/1367-2630/aa8b9b} {\bibfield  {journal} {\bibinfo
  {journal} {New J. Phys.}\ }\textbf {\bibinfo {volume} {19}},\ \bibinfo
  {pages} {113020} (\bibinfo {year} {2017}{\natexlab{b}})}\BibitemShut
  {NoStop}%
\bibitem [{\citenamefont {Tavakoli}\ \emph {et~al.}(2017)\citenamefont
  {Tavakoli}, \citenamefont {Renou}, \citenamefont {Gisin},\ and\ \citenamefont
  {Brunner}}]{tavakoli2017correlations}%
  \BibitemOpen
  \bibfield  {author} {\bibinfo {author} {\bibfnamefont {A.}~\bibnamefont
  {Tavakoli}}, \bibinfo {author} {\bibfnamefont {M.-O.}\ \bibnamefont {Renou}},
  \bibinfo {author} {\bibfnamefont {N.}~\bibnamefont {Gisin}},\ and\ \bibinfo
  {author} {\bibfnamefont {N.}~\bibnamefont {Brunner}},\ }\href
  {https://doi.org/10.1088/1367-2630/aa7673} {\bibfield  {journal} {\bibinfo
  {journal} {New J. Phys.}\ }\textbf {\bibinfo {volume} {19}},\ \bibinfo
  {pages} {073003} (\bibinfo {year} {2017})}\BibitemShut {NoStop}%
\bibitem [{\citenamefont {Wolfe}\ \emph {et~al.}(2019)\citenamefont {Wolfe},
  \citenamefont {Spekkens},\ and\ \citenamefont {Fritz}}]{wolfe2019inflation}%
  \BibitemOpen
  \bibfield  {author} {\bibinfo {author} {\bibfnamefont {E.}~\bibnamefont
  {Wolfe}}, \bibinfo {author} {\bibfnamefont {R.~W.}\ \bibnamefont
  {Spekkens}},\ and\ \bibinfo {author} {\bibfnamefont {T.}~\bibnamefont
  {Fritz}},\ }\bibfield  {journal} {\bibinfo  {journal} {J. Causal Inference}\
  }\textbf {\bibinfo {volume} {7}},\ \href
  {https://doi.org/doi:10.1515/jci-2017-0020} {doi:10.1515/jci-2017-0020}
  (\bibinfo {year} {2019})\BibitemShut {NoStop}%
\bibitem [{\citenamefont {Wolfe}\ \emph {et~al.}(2021)\citenamefont {Wolfe},
  \citenamefont {Pozas-Kerstjens}, \citenamefont {Grinberg}, \citenamefont
  {Rosset}, \citenamefont {Ac\'{\i}n},\ and\ \citenamefont
  {Navascu\'es}}]{wolfe2021qinflation}%
  \BibitemOpen
  \bibfield  {author} {\bibinfo {author} {\bibfnamefont {E.}~\bibnamefont
  {Wolfe}}, \bibinfo {author} {\bibfnamefont {A.}~\bibnamefont
  {Pozas-Kerstjens}}, \bibinfo {author} {\bibfnamefont {M.}~\bibnamefont
  {Grinberg}}, \bibinfo {author} {\bibfnamefont {D.}~\bibnamefont {Rosset}},
  \bibinfo {author} {\bibfnamefont {A.}~\bibnamefont {Ac\'{\i}n}},\ and\
  \bibinfo {author} {\bibfnamefont {M.}~\bibnamefont {Navascu\'es}},\ }\href
  {https://doi.org/10.1103/PhysRevX.11.021043} {\bibfield  {journal} {\bibinfo
  {journal} {Phys. Rev. X}\ }\textbf {\bibinfo {volume} {11}},\ \bibinfo
  {pages} {021043} (\bibinfo {year} {2021})}\BibitemShut {NoStop}%
\bibitem [{\citenamefont {Jackson}(1999)}]{jackson1999classical}%
  \BibitemOpen
  \bibfield  {author} {\bibinfo {author} {\bibfnamefont {J.~D.}\ \bibnamefont
  {Jackson}},\ }\href@noop {} {\emph {\bibinfo {title} {Classical
  electrodynamics}}}\ (\bibinfo  {publisher} {American Association of Physics
  Teachers},\ \bibinfo {year} {1999})\ pp.\ \bibinfo {pages}
  {527--529}\BibitemShut {NoStop}%
\end{thebibliography}%

\end{document}